\documentclass[conference]{IEEEtran}
\usepackage{doublespace}
\IEEEoverridecommandlockouts

\usepackage[cmex10]{amsmath}
\usepackage{centernot}
\usepackage{array}
\usepackage{url}

\usepackage{graphicx}
\usepackage{amsmath} 
\usepackage{cite}

	\newtheorem{theorem}{Theorem}

\newenvironment{example}{\par\begin{quotation}\small\noindent{\bf Example:\ }}{\end{quotation}\par}

\newenvironment{Proof}[1]{\medskip\par\noindent 
{\bf Proof:\, #1}\,}{{\mbox{\,$\bullet$}\par}}

	\newcommand{\thmref}[1]{(\ref{thm:#1})}
	\newcommand{\Thmref}[1]{Theorem~\ref{thm:#1}}
	\newcommand{\thmlabel}[1]{\label{thm:#1}}

	\newcommand{\twodef}[4]{\left\{\begin{array}{ll} 
		\displaystyle {#1} & {#2} \\
		\displaystyle {#3} & {#4}
		\end{array}\right.}

\newcommand{\imc}{{inscribed matter}}
\newcommand{\IMC}{{Inscribed Matter}}

\newcommand{\bG}{{\bar{G}}}

\newcommand{\blankout}[1]{}

\newcommand{\note}[1]{{\bf (NOTE:  #1 ) }}
\newcommand{\be}{\begin{equation}}
\newcommand{\ee}{\end{equation}}
\newcommand{\bee}{\begin{IEEEeqnarray}{c}}
\newcommand{\eee}{\end{IEEEeqnarray}}
\newcommand{\bs}{\begin{IEEEeqnarray*}{c}}
\newcommand{\es}{\end{IEEEeqnarray*}}
\newcommand{\bd}{\begin{IEEEeqnarray*}{c}}
\newcommand{\ed}{\end{IEEEeqnarray*}}
\newcommand{\bea}{\begin{IEEEeqnarray}{rCl}}
\newcommand{\eea}{\end{IEEEeqnarray}}
\newcommand{\bas}{\begin{IEEEeqnarray*}{rCl}}
\newcommand{\eas}{\end{IEEEeqnarray*}}
\newcommand{\equat}[1]{equation (\ref{eq:#1})}
\newcommand{\Equat}[1]{Equation (\ref{eq:#1})}

\newcommand{\xv}{{\bf x}}
\newcommand{\sv}{{\bf s}}

\newcommand{\gv}{{\mathbf{g}}}

\newcommand{\tv}{{\bf t}}

\newcommand{\uv}{{\bf u}}

\newcommand{\Hup}{{H^{\uparrow}}}
\newcommand{\tvv}{\vec{\tv}}
\newcommand{\tvec}{\vec{t}}

\newcommand{\svec}{\vec{s}}
\newcommand{\Svec}{\vec{S}}
\newcommand{\svv}{\vec{\sv}}
\newcommand{\Sv}{\vec{S}}
\newcommand{\Svv}{\vec{\Smat}}
\newcommand{\Tvv}{\vec{\Tmat}}
\newcommand{\Xvv}{\vec{\Xmat}}

\newcommand{\Pmat}{{\bf P}}

\newcommand{\bzero}{{\bf 0}}
\newcommand{\binfty}{{\boldsymbol{\infty}}}

\newcommand{\Dmat}{{\bf D}}

\newcommand{\Tmat}{{\bf T}}
\newcommand{\Smat}{{\bf S}}
\newcommand{\Xmat}{{\bf X}}

\newcommand{\na}{{\Omega}}

\newcommand{\mymu}{{\mu}}
\newcommand{\mylambda}{{\lambda}}

\newcommand{\myrho}{{\rho}}

\newcommand{\btau}{{\boldsymbol{\tau}}}

\author{Christopher Rose~\IEEEmembership{Fellow,~IEEE} and I.S. Mian}

\title{Inscribed Matter Communication: Part I}

\begin{document}

\maketitle

\begin{abstract}
We provide a fundamental treatment of the molecular communication channel wherein ``inscribed
matter'' is transmitted across a spatial gap to provide reliable signaling between a sender and
receiver. Inscribed matter is defined as an ensemble of ``tokens'' (biotic/abiotic objects) and is
inspired, at least partially, by biological systems where groups of individually constructed
discrete particles ranging from molecules to viruses and organisms are released by a source and
travel to a target -- for example, morphogens or semiochemicals diffuse from one cell, tissue or
organism to another.  For identical tokens that are neither lost nor modified, we consider messages
encoded using three candidate communication schemes: a) token timing (timed release), b) token
payload (composition), and c) token timing plus payload. We provide capacity bounds for each scheme
and discuss their relative utility. We find that under not unreasonable assumptions, megabit per
second rates could be supported at $100$ femtoWatt transmitter powers. Also, since quantities such as token
concentration or token-counting are derivatives of token arrival timing, token timing
undergirds all molecular communication techniques. Thus, our modeling and results about the physics
of efficient token-based information transfer can inform investigations of diverse theoretical and
practical problems in engineering and biology. This work, Part I, focuses on the information
theoretic bounds on capacity.  Part-II develops some of the mathematical and information-theoretic
machinery that support the bounds presented here.
\end{abstract}

\section{Introduction}

Scale-appropriate signaling methods become important as systems shrink to the
nanoscale. For systems with feature sizes of microns and smaller, electromagnetic
and acoustic communication become increasingly inefficient because energy coupling
from the transmitter to the medium and from the medium to the receiver becomes
difficult at usable frequencies. Biological systems, with the benefit of lengthy
evolutionary experimentation, seem to have arrived at a ubiquitous solution to
this signaling problem at small and not so small scales: use of ``inscribed matter''
(an ensemble of discrete particles) which travels through some material bearing
a message from one entity to another. Broad classes of such particles include
\begin{itemize}
\item
{\em Molecules} such as electronically activated species, ions, chemicals,
biopolymers, and macromolecular complexes.
\item
{\em Membrane-bound structures} such as intra- and extracellular vesicles (for
instance, exosomes, microvesicles, apoptotic bodies, ectosomes, endosomes, lysosomes,
autophagosomes, and vacuoles) and intracellular organelles (for instance, nuclei,
mitochondria, and chloroplasts).
\item
{\em Cells} such as stem cells, tumor cells, and hematocytes.
\item
{\em Acellular, unicellular and multicellular life forms} (organisms for brevity)
such as viruses, viroids, phages, plasmids, bacteria, archaea, fungi, protists,
plants, and animals.
\item
{\em Objects} such as matter in the natural world (for instance pollen grains,
seeds, and proteinaceous aggregates such as prions); and human artifacts (for
example, Voyager Golden Records).
\end{itemize}

Studies of engineered nano-scale communication systems have focused on the
encoding, transmission, and decoding of information using patterns of one
category of discrete particles, namely molecules. A large portion of this
work in ``molecular communication'' has considered time-varying
concentration profiles of molecules as the fundamental signal measurement
\cite{fekriisit11,akyildiz_nanonet,akyildiz_diffusion,Eckbook,ISIT2013_Arash,arash_wireless2013,ICC2014_Arash}.
However, concentration is a {\em collective} property of the process and
masks the underlying physics of molecule release by the sender and capture
by the receiver. This begs the questions of truly fundamental limits for
communication using ensembles of molecules in particular, discrete particles
more broadly, and what we term ``tokens'' in general.

This paper is organized as follows:

First, we discuss communication using inscribed matter from biological and engineering
perspectives. We illustrate how scenarios spanning a wide range of spatial and temporal scales and
from seemingly disparate disciplines can be understood within a unified framework: the token timing
and/or token payload channel, a communication scheme wherein information is carried from sender to
receiver by tokens via their timed release, their composition, or both.We will assume tokens always
(eventually) arrive, and are removed from circulation upon first seizure by the receiver.  This
abstraction encompasses not only token timing but also the token concentration and token counting
models prevalent in the molecular communication literature
\cite{fekriisit11,akyildiz_nanonet,akyildiz_diffusion,Eckbook,ISIT2013_Arash,arash_wireless2013,ICC2014_Arash,farsad_isit16,farsadIT16}.

Next, we describe the token timing channel wherein information is encoded {\em only} in the release
time of {\em identical} tokens as opposed to inscribed onto tokens (tokens with payloads) or in the
number of tokens released (token counting).  Though seemingly limited, this pure timing model
supplies the mathematical machinery to precisely consider both token payload and token counting
communication schemes.  To this end, we provide a mathematical formulation of token timing channel:
identical tokens emitted with independent stochastic (but asymptotically assured, one-time)
arrivals. We formalize the signaling model so that the typical energy-dependent asymptotic sequential
channel use coding results based on mutual information between input and output can be applied
\cite[(chapt 8 \& 10)]{cover}. We then show how these results can be applied to token counting and
tokens with payloads.  We focus on molecular tokens -- particularly DNA and protein sequences since
their energy requirements (and information content) are well-understood -- and show that information
transfer using {\imc} can be extremely efficient.
We find that megabit per second rates could be supported theoretically with
on the order of $100$ femtoWatts of transmitter power.

Finally, we explore how our studies and the attendant insights could aid biological
understanding of and inform engineering approaches to {\imc} communication.

\section{Inscribed Matter}

\subsection{Communication using discrete particles: a natural world perspective}

Networks of intercommunicating biological entities occur at whatever level one
cares to consider: (macro)molecules, cells, tissues, organisms, populations,
microbiomes, ecosystems, and so on. An ancient yet still widespread method for
one entity to convey a message to another is via inscribed matter. The typical
scenario is as follows: information-bearing discrete particles are released by
a source, travel through a material, and are captured by a target where they
are interpreted. The following examples illustrate the diversity and complexity
of such inscribed matter communication (the particles are italicized).
\begin{itemize}
\item
{\em Electrons} from an electron donor flow through an electron transport chain
to an electron acceptor where the electrochemical gradient is used convert
mechanical work into chemical energy as part of a cellular process such as
photosynthesis or respiration. In microbial communities, electrons are
transferred from one individual to another through bacterial nanowires
(electrically conductive appendages), bacterial cables (thousands of individuals
lined up end-to-end with electron donors located in the deeper regions of marine
sediment and electron acceptors positioned in its upper layers where oxygen is
more abundant), and biofilms (community members embedded in a self-produced
three-dimensional matrix of extrapolymeric substances) \cite{ubli15}.
\item
{\em Free radicals} produced from molecules in the nucleoplasm by the direct or
indirect action of ionizing radiation diffuse to the genome where they alter/damage
nucleotide bases and sugars.
\item
{\em messenger RNA (mRNA) molecules} transcribed from a eukaryotic genome in a
nucleus migrate to ribosomes in the cytoplasm where they are translated into
proteins.
\item
{\em Acetylcholine (ACh) molecules} released by a vertebrate motor neuron diffuse
across the synapse to nicotinic ACh receptors on the plasma membrane of the muscle
fiber where binding triggers muscular contraction.
\item
{\em Homing endonuclease (HE) containing inteins} self-excised from bacterial,
archaeal, eukaryotic or viral host proteins home to a target site in the genome
of the same or different organism where the genetic parasitic element reinserts
itself into the intein-free allele of the host gene (horizontal dissemination);
inteins without a functioning HE are mainly transferred vertically but may move
horizontally along with the host gene.
\item
{\em Ions, molecules, organelles, bacteria and viruses} present in one cell travel
through a thin membrane channel (tunneling nanotube) to the physically connected
cell where they elicit a response.
\item
{\em Semiochemicals} (chemical substances or mixtures of volatile molecules)
emitted by one individual travel to another of the same (pheromones) or different
species (allelochemicals) where they elicit a response -- allomones benefit only
the sender, kairomone benefit only the receiver, and synomones benefit both.
\item
{\em Extracellular vesicles} secreted by all living cells -- including bacteria,
archaea and eukaryotes -- and harboring specific cargo materials (for instance,
proteins, nucleic acids, lipids, metabolites, antigens, and viruses) traverse the
extracellular space or body fluids (for instance, blood and urine) to a local or
distal recipient cell where they transfer their bioactive contents.
\item
{\em Cargo-bearing molecular motors} shuttle along a track system of cytoskeletal
filaments to another point in the cell compartment where their freight such as
vesicles containing molecules and tubes is unloaded.
\item
{\em Single and clusters of metastatic cells} that have escaped from a primary
tumor circulate through the blood or lymph to a secondary organ site where, after
extravasation, they can seed a new tumor.
\item
{\em Organic particles} such as microorganisms, fungal spores, small insects, and pollen grains
associated with a macroorganism, geological site or geographic location relocate to another host or
region where they influence the local biochemistry, geochemistry and climate \cite{MorSan12}-- long
distance transport (including movement within and between continents and oceans) can occur via the
same meteorological phenomena and processes, such as jetstreams and hurricanes, that translocate
non-biological particles such as sea salt and dust.
\item
{\em Crustal material} ejected by a Solar System body travels to another body
where if it carries microbial spores or building blocks such as amino acids,
nucleobases and lipid-like molecules has the potential to seed life.
Ejecta (potentially carrying microbial spores) travel from Ceres (the largest
object in the asteroid belt which lies between the orbits of Mars and Jupiter)
to terrestrial planets in the solar system (Earth, Mars or Venus) \cite{Hou11};
the presence of water on Ceres \cite{KupORo14} suggests the dwarf planet's
potential as a home for extraterrestrial life.
\end{itemize}

Irrespective of the precise nature of the components of the inscribed matter communication system --
the discrete particles (information carriers), source (sender), spatial gap (transmission medium),
and target (receiver) -- two fundamental questions are ``How reliable is communication?,'' and ``How
is useful information conveyed given constraints on resources?.''  Here, we investigate token timing
(discrete particle release and capture times) and token payload (energy required to manufacture
discrete particles, to assemble symbolic strings from a set of building blocks -- we do not consider
the energy required for de novo synthesis of the building blocks).  And although not explicitly
stated, please note that our energy model could also include token sequestration, token ejection,
and token transport, the active movement of discrete particles (the energetics of translocating
vesicles by a molecular motor system which converts chemical or other form of energy into mechanical
energy).  The only requirement is that the energy cost per token is independent of the information
carried.

In the token timing channel model we will elaborate later on, tokens are neither lost nor modified:
the number and makeup of the tokens emitted by the source is the same as the ones arriving at the
target, all that differs are their times of emission and their times of arrival. While accommodating
tokens that are delayed temporarily, our mathematical model does not directly consider tokens that
are detained permanently, removed entirely, never arrive, or are changed en route. In the natural
world, discrete particles often interact with the material through which they travel resulting in
their immurement and ultimate removal or detention and eventual discharge. Examples include:
\begin{itemize}
\item
{\em Free radicals} produced by radiolysis may react chemically with neighboring
materials.
\item
{\em mRNAs} may be modified post-transcriptionally.
\item
{\em ACh} can be degraded by the enzyme acetylcholine esterase present in synapses.
\item
The random path of a {\em semiochemical} diffusing through air, soil or water may
result in a trajectory that leads away from the destined individual.
\item
{\em Circulating tumor cells} may be destroyed by the immune system.
\item
{\em Microscopic particles} may be immobilized within mucus -- the polymer-based
hydrogel covering the inner linings of the body -- depending on the density
of the mucin network and environmental factors such as pH and ionic strength.
\item
{\em Bacteria}, particularly plant pathogens, present in the atmosphere can nucleate
the formation of ice in clouds resulting in snow, rain and hail \cite{MorSan12}.
\end{itemize}

Nonetheless, our model does provide an organizing principle for all forms of molecular
communications since these sorts of impediments -- token loss or corruption -- can only decrease the
capacity of the system we analyze.  Furthermore, the analysis is ``compartmental'' in the sense that
token corruption and loss can be treated separately without invalidating the fundamental ``outer
bound'' results.

\subsection{Communication using physical objects: an engineering perspective}
\label{sect:bioperspective}

Inscribed matter can often be the most energy-efficient means of communication when
delay can be tolerated. In fact, a once popular communication networks textbook
\cite{tannenbaum} contains the passage:
\begin{quote}
{\em Never underestimate the bandwidth of a station wagon full of tapes hurtling down the highway.

\hfill  {\small -- A.S. Tanenbaum, {\em Computer Networks}, 4th ed., p. 91}
}
\end{quote}
This somewhat tongue-in-cheek ``folklore'' should come as no surprise. From
early antiquity, private persons, governments, the military, press agencies,
stockbrokers and others have used carrier pigeons to convey messages.
Today, ``sneakernets'' \cite{microsoft} have been proposed as a low-latency high-fidelity network
architecture for quantum computing across global distances: ships carry
error-corrected quantum memories installed in cargo containers \cite{DevGre14}.

Previous work on mobile wireless communication found that network capacity could be increased if
delay-tolerant traffic was queued until the receiver and sender were close to one another -- perhaps
close enough to exchange physical storage media \cite{Frenkiel,wcnc12_rose,
  infostations-vtc,irvine,infostation1,dect,infostations-pc,ana,alap_book,ana-thesis,iacono,mineminemine,demonfuruzan,furuzan_ciss}.
This recognition prompted a careful consideration of the energetics involved in delivery of physical
messages, and a series of papers \cite{rose_IM_asilomar,rose_IM,rose_wright_nature} revealed the
surprising results that {\imc} can be many orders of magnitude more efficient than radiative methods
even with moderate delay constraints and over a variety of size scales. In fact,
\cite{rose_wright_nature} showed that over interstellar distances (10k light years), {\imc} could be
on the order of $10^{15}$ times more energy-efficient than radiated messages, suggesting that
evidence of extraterrestrial civilizations would more likely come from artifacts than from radio
messages if energy requirements are a proxy for engineering difficulty \cite{NSF_ET_Discoveries}.

At the other end of the size scale there has been increasing interest in
biologically-inspired {\imc} communication at the nano/microscale \cite{JSACmesocom}
where the information carrier ranges from timed release of identical signaling agents
to specially constructed information carriers \cite{bassler1999,bassler2002,fekriisit11,akyildiz_nanonet,akyildiz_diffusion,eckford1,Eckbook,ISIT2013_Arash,arash_wireless2013,ICC2014_Arash,eckford_press_vodka}.
Although this field of molecular communication is in its infancy with seemingly
futuristic application plans currently out of reach (for instance, {\em in vivo}
biological signaling, surgical/medicinal/environmental microbot swarms, or process-on-a-chip),
the theoretical potential rates and energy efficiencies, especially through media
unfriendly to radiation, are sufficiently large \cite{icc15} to warrant careful
theoretical and practical consideration.

\begin{figure}
\begin{center}
\vspace{-0.25in}
\includegraphics[height=0.85in,width=2.5in]{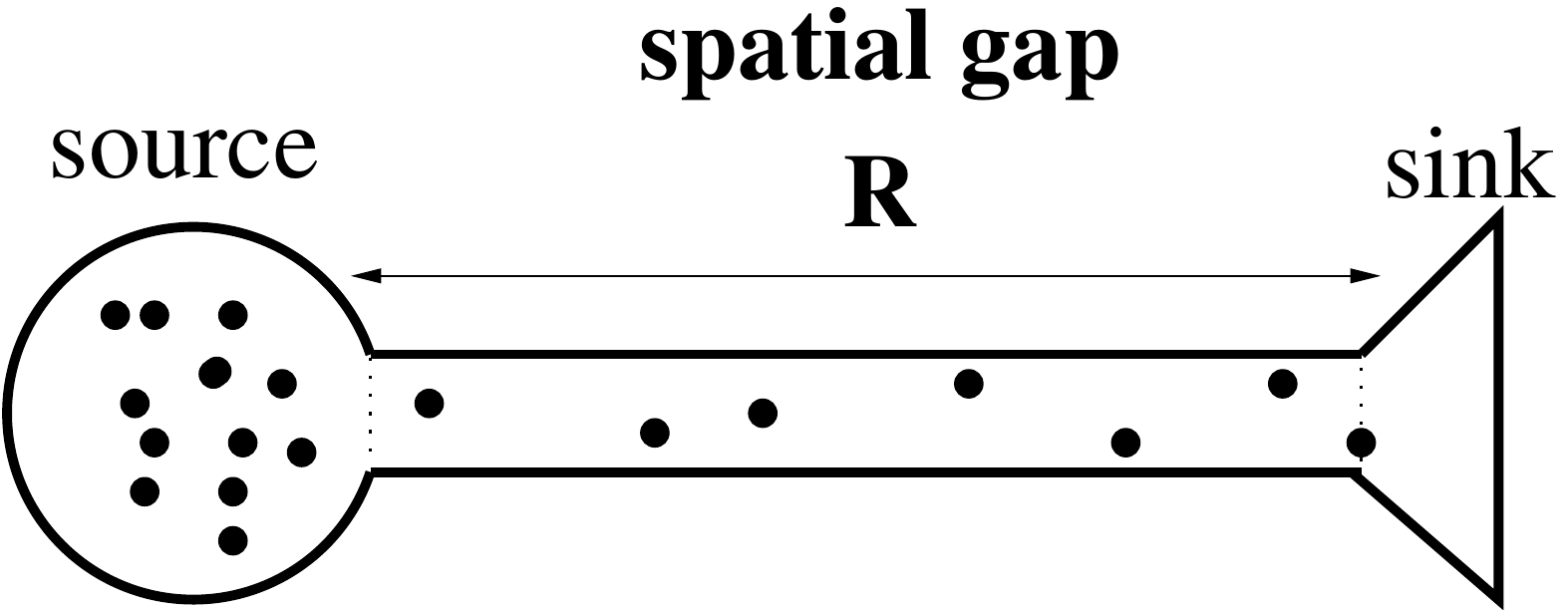}
\includegraphics[height=1.5in,width=2.5in]{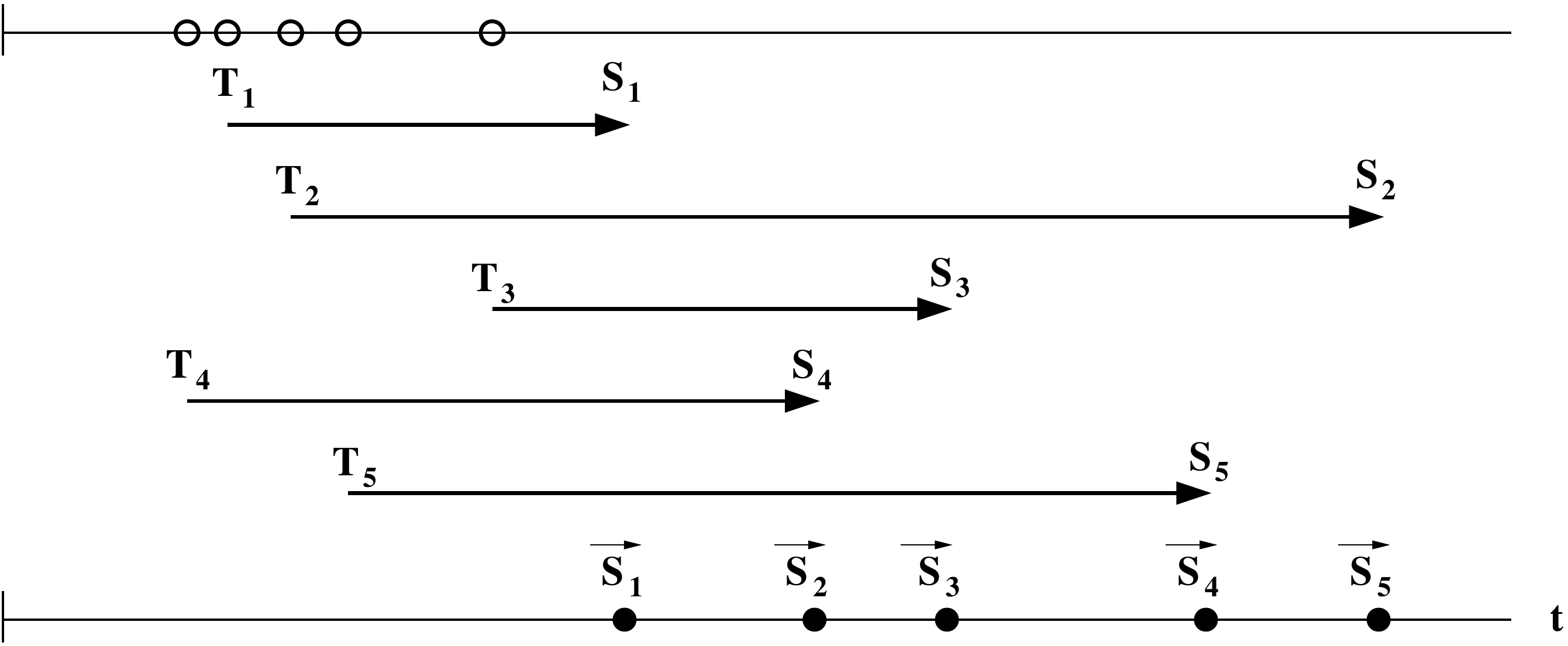}
\end{center}
\caption{An Abstraction of an {\IMC} Communication Channel.
A sender transmits an ensemble of tokens (``inscribed matter'') to a
receiver across a spatial gap (of length $R$ in the figure). The tokens are released at (unordered) times
$\{T_m\}$, propagate through a transmission medium and are captured at
corresponding times $\{S_m\}$. For identical tokens, the receiver sees ordered
arrivals $\{ \Svec_m \}$ which may differ in index from the unordered arrivals
$\{ S_m\}$.}
\label{figure:imchannel}
\end{figure}

From an engineering perspective, the basic idea of {\imc} communication is very simple
(FIGURE~\ref{figure:imchannel}). Information is coded in the structure of the signaling agent and/or
its release time at the sender. These agents traverse some spatial gap to the receiver where they
are captured and the information decoded.  There are, of course, many details and variations on the
theme. As explicitly mentioned for biological systems, the signaling agents (or tokens as we call
them) could be identical, implying that timing (which includes time-varying concentration) is {\em
  \bf the only} information carrier, or tokens could themselves carry data payloads (in addition to,
or in lieu of timing).  However, unlike the award-winning paper ``Bits Through Queues'' by
Anantharam and Verd\'u \cite{bits-Qs} and later by Sundaresan and Verd\'u \cite{sundaresan1,
  sundaresan2}, we do not know which arrival times correspond to which emission times.

The ``gap'' (channel) could be a medium through which tokens diffuse stochastically, or some form of
active transport might be employed.  In addition, tokens could be deliberately ``eaten'' by
gettering agents injected by the sender, channel or receiver.
Similarly, tokens could be corrupted during passage through the channel or might
simply get ``lost'' and never reach the receiver \cite{eckford1,farsadIT16}.

Furthermore, the reception process itself could be noisy. If we sought to mimic biological systems,
a typical receptor structure is stereochemically matched to a particular signaling molecule (token)
and the kinetics of the ligand binding/unbinding process must be considered as well as the number
and density of receptors. Furthermore, a given receptor may preferentially bind to a ligand (token),
but there may be other different or identical (but from another source) {\em interfering} ligands
which bind to the same receptor. When one considers networks of molecular transceivers, this sort of
``cross talk'' or outright interference must be considered.

\subsection{Inscribed matter communication: model distillation}

While the various engineering and biological scenarios require slightly different
information theoretic formulations, they can {\em all} be understood within a unified
framework: the identical token timing channel wherein
\begin{itemize}
\item
Token release and capture timing is the only mechanism for information transfer.
\item
Tokens always (eventually) arrive at the receiver.
\item
Tokens are removed promptly from circulation (or deactivated) after first reception.
\end{itemize}

The identical token timing channel abstraction \cite{wcnc12_rose, SonRos13,isit13,isit14,icc15,farsad_isit16}
encompasses token concentration or token counting models since time-varying concentration (or token
counts in ``bit intervals'') at a receiver is a coarse-time approximation to the precise individual
token timing model.

The timing channel is also important for understanding information carriage via payload-charged
tokens whose information packets may need resequencing at the receiver. That is, timing channel
results provide tight bounds on resequencing overhead and are especially important if it is
technologically difficult to construct tokens with large payloads.  In addition, the operation of
the timing channel also sheds light on channels wherein the number of tokens sent is the information
carrier during signaling intervals \cite{farsad_isit16}.

In addition, as mentioned at the end of section~\ref{sect:bioperspective}, the timing channel
provides {\em outer bounds} since the uncertainty associated with various receptor and channel
models can only {\em decrease} the information-carrying capacity of the channel via the data
processing theorem \cite[(chapt 2)]{cover}. For instance, re-capture processes owing to receptor
binding kinetics \cite{eckford1} and token loss (erasure) can only decrease the channel
capacity. Likewise, token processing/corruption/loss can again only decrease channel capacity. Thus,
the timing channel not only allows upper limits on capacity to be obtained, but also permits the
overall channel to be treated as a cascade, each constituent of which can be analyzed separately and
compared to identify potential information transfer bottlenecks.

\section{Mathematical Formulation}
TABLE~\ref{table:glossary} is a glossary of key quantities that will be discussed in what follows.  For
continuity and clarity, the identical table is included in the companion paper, Part-II
\cite{RoseMian16_2}.
\begin{table}[h]
\begin{tabular}{p{2.0cm}|p{5.9cm}}  \hline
{\bf \em Token} &  {\small A unit released by the transmitter and captured by the receiver}\\ \hline
{\bf \em Payload} & {\small Physical information (\imc) carried by a token}\\ \hline
{\bf \em $\mylambda$} & {\small The average rate at which tokens are released/launched into the channel}\\ \hline
{\bf $\Tmat$} & {\small A vector of token release/launch times}\\  \hline
{\bf \em First-Passage} & {\small The time between token release/launch and token capture at the receiver}\\ \hline
{\bf $\Dmat$} & {\small A vector of first-passage times associated with launch times $\Tmat$}\\  \hline
{\bf $G(\cdot)$} & {\small The cumulative distribution function for first-passage random variable $D$}\\  \hline
{\bf $1/\mymu$} & {\small Average/mean first-passage time}\\  \hline
{\bf $\myrho$} & {\small $\mylambda/\mymu$, a measure of system token ``load'' }\\  \hline
{\bf $\Smat$} & {\small A vector of token arrival times, $\Smat = \Tmat + \Dmat$} \\  \hline
{\bf $P_k(\xv)$} & {\small A permutation operator which rearranges the order of elements in vector $\xv$}\\ \hline
{\bf $\Omega$} & {\small The ``sorting index'' which produces $\Svv$ from $\Smat$, {\em i.e.}, $\Svv = P_{\Omega}(\Smat)$}\\ \hline
{\bf $\Svv$} & {\small An {\em ordered} vector of arrival times obtained by sorting the elements of $\Smat$ (note, the receiver only sees $\Svv$ not $\Smat$)}\\  \hline
{\bf $I(\Smat;\Tmat)$} & {\small The mutual information between the launch times (input) and the arrival times (output)}\\  \hline
{\bf $I(\Svv;\Tmat)$} & {\small The mutual information between the launch times (input) and the {\em ordered} arrival times (output)}\\  \hline
{\bf $h(\Smat)$} & {\small The differential entropy of the arrival vector $\Smat$}\\  \hline
{\bf $H(\Omega|\Svv,\Tmat)$} & {\small The {\em ordering entropy} given the input $\Tmat$ and the output $\Svv$} \\  \hline
{\bf $\Hup(\Tmat)$} & {\small An upper bound for $H(\Omega|\Svv,\Tmat)$}\\  \hline
{\bf $C_q$ {\rm and } $C_t$} & {\small The asymptotic per token and per unit time capacity between input and output}\\  \hline
  \end{tabular}
$\mbox{ }$\\
\caption{Glossary of useful terms}
\label{table:glossary}
\end{table}

\begin{figure}
\begin{center}
\vspace{-0.5in}
\includegraphics[height=2.0in,width=2.0in]{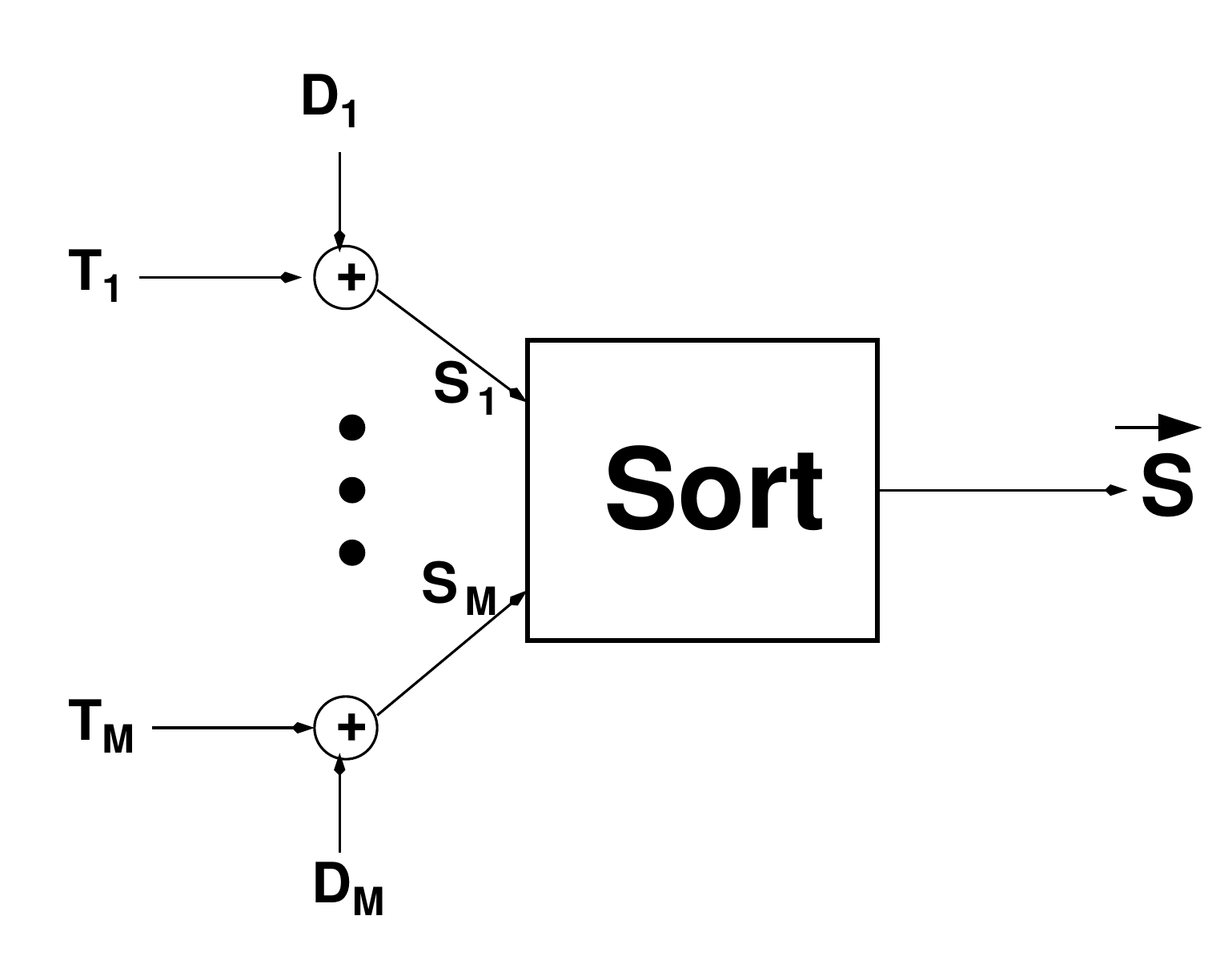}
\end{center}
\caption{The token release with reordering {\IMC} communication channel.  For token $m$
  released at time $T_m$, the duration of its first-passage between the sender and receiver is $D_m$
  so it arrives at time $S_m$.  The $\{S_m \}$ are then sorted by order of arrival. Since the $M$
  tokens are identical, the ordered arrival time $\Svec_m$ may not correspond to $S_m$.  }
\label{figure:channel}
\end{figure}

Following \cite{isit11,wcnc12_rose, isit13, isit14, icc15}, assume emission of $M$ identical tokens
at times $\{T_m\}$, and their capture at times $\{S_m \}$, $m = 1,2,\cdots,M$.  The duration of token
$m$'s first-passage between source and destination is $D_m$. These $D_m$ are assumed i.i.d. with
$f_{D_m}(d) = g(d) = G^{\prime}(d)$ where $g(\cdot)$ is some causal probability density with mean
$\frac{1}{\mymu}$ and cumulative distribution function (CDF) $G(\cdot)$.  We also assume that $g(\cdot)$
contains no singularities.  Thus, the first portion of the channel is modeled as a sum of random
$M$-vectors
\bee
\Smat = \Tmat + \Dmat
\eee
as shown in FIGURE~\ref{figure:channel} prior to the sorting operation.

We therefore have
\bea
\label{eq:f_s_def}
f_{\Smat}(\sv) & = & 
{ \int_{\bzero}^{\binfty}}  f_{\Tmat}(\tv) f_{\Smat|\Tmat}(\sv|\tv)  d \tv \nonumber\\
 & = & 
{\int_{\bzero}^{\sv}} f_{\Tmat}(\tv)  \prod_{m=1}^M g(s_m - t_m)  d \tv   \nonumber\\
 & = & 
{\int_{\bzero}^{\sv}} f_{\Tmat}(\tv)  \gv(\sv - \tv)  d \tv 
\eea
where
\bd
\gv(\sv - \tv)
=
\prod_{m=1}^M g(s_m - t_m) 
\ed
We impose an emission deadline, $T_m \le \tau$, $\forall m \in \{ 1,2,\cdots,M \}$.  The associated
emission time ensemble probability density $f_{\Tmat}(\tv)$ is assumed causal, but otherwise
arbitrary.  Had we imposed a mean constraint instead of a deadline, the channel between $\Tmat$ and
$\Smat$ would be the parallel version of Anantharam and Verd\'u's {\em Bits Through Queues}
\cite{bits-Qs}.  However, since the tokens are identical we cannot necessarily determine which
arrival corresponds to which emission time.  That is, the final output of the channel is a reordering
of the $\{ s_m \}$ to obtain a set $\{ \vec{s}_m \}$ where $\vec{s}_m \le \vec{s}_{m+1}$,
$m=1,2,\cdots, M-1$, as shown on the right hand side of FIGURE~\ref{figure:channel} after the
sorting operation.

We write this relationship as
\bee
\label{eq:permutationdef}
\Svv = P_{\na}(\Smat)
\eee
where $P_{\na}(\cdot)$, ${\na}=1,2,\cdots,M!$, is a permutation operator and $\na$ is that permutation index which
produces   ordered $\Svv$ from the argument $\Smat$. 
We define $P_1(\cdot)$ as the identity permutation operator, $P_1(\sv) = \sv$.

We note that the event $S_i = S_j$ ($i \ne j$) is of zero measure owing to the
no-singularity assumption on $g(\cdot)$, Thus, for analytic convenience we will assume
that $f_{\Smat}(\sv) = 0$ whenever two or more of the $s_m$ are equal and
therefore that the $\{ \vec{s}_m \}$ are strictly ordered wherever $f_{\Svv}(\cdot) \ne 0$
({\em i.e.}, $\vec{s}_m < \vec{s}_{m+1}$).

Thus, the density $f_{\Svv}(\svv)$ can be found by ``folding'' the density
$f_{\Smat}(\sv)$ about the hyperplanes described by one or more of the $s_m$
equal until the resulting probability density is nonzero only on the region
where $s_m < s_{m+1}$, $m=1,2,\cdots,M-1$.  Analytically we have
\bee
\label{eq:svv_def}
f_{\Svv}(\svv)
=
\sum_{n=1}^{M!}
f_{\Smat}(P_{n}(\svv))
\eee
Then, since $f_{S|T}(s|t) = g(s-t)$, we can likewise describe $f_{\Svv|\Tmat}(\sv|\tv)$ as
\bee
\label{eq:svvT_def}
f_{\Svv|\Tmat}(\sv|\tv)
=
\sum_{n=1}^{M!}
\gv(P_n(\sv) - \tv)
\prod_{m=1}^M
u([P_n(\sv)]_m - t_m)
\eee
again for ${s_1 < s_2 < \cdots < s_m}$ and zero otherwise.
With exponential first-passage, $g(d) = \mymu e^{-\mymu d}u(d)$, becomes
\bee
\label{eq:econd}
f_{\Svv|\Tmat}(\sv|\tv)
=
\mymu^M
e^{ - \mymu \displaystyle{\sum_{i=1}^M} (s_i - t_i)}
\left (
\sum_{n=1}^{M!}
\uv(P_n(\sv) - \tv)
\right )
\eee
again assuming $s_1 < s_2 < \cdots < s_m$.  It is worth mentioning explicitly that \equat{econd}
{\em does not assume} arguments $s_i \ge t_i$ as might be implicit in \equat{f_s_def}.

Finally, the problem structure will allow us to make use of multi-dimensional function symmetry
(hypersymmetry) arguments, $f(\xv) = f(P_n(\xv))$ $\forall$ permutations $n$.  The following
property of expectations of hypersymmetric functions over
hypersymmetric random variables will later prove useful.
{\em   \begin{theorem} {\bf Hypersymmetric Expectation:}
\thmlabel{Ehypersymmetry}

Suppose $Q(\xv)$ is a hypersymmetric function, $Q(\xv) = Q(P_k(\xv))$ $\forall k$, and $\Xmat$ is a
hypersymmetric random vector.  Then, when $\vec{\Xmat}$ is the ordered version of random vector
$\Xmat$ we have
\bee
\label{eq:hyperexpect}
E_{\Xvv} \left [Q(\Xvv) \right ]
=
E_{\Xmat} \left [Q(\Xmat) \right ]
\eee
\end{theorem}}
\begin{Proof}{\Thmref{Ehypersymmetry}}
$\tilde{\Xmat}$ is a deterministic function of $\Xmat$; i.e., $\theta(\Xmat) = {\tilde{\Xmat}}$.  Thus,
\bee
E \left [
Q(\tilde{\Xmat})
  \right ]
=
E \left [
Q(\theta(\Xmat))
  \right ]
=
E \left [
Q(\Xmat)
  \right ]
\eee
where the last equality results from the hypersymmetry of $Q(\cdot)$.
\end{Proof}
With these preliminaries done, we can now begin to examine the mutual information between the
unordered emission times $\Tmat$, the unordered arrival times $\Smat$, and the ordered (sorted)
arrival times $\Svv$.

\section{Information Theoretic Analysis}

\subsection{Formalizing The Signaling Model}
\label{sect:channeluse}
To determine whether the mutual information between $\Tmat$ the input and $\Svv$ the output is a
measure of channel capacity, it suffices to have a signaling model which patently supports the usual
asymptotically large block length and repeated independent sequential channel uses paradigm \cite[(chapt 8 \&
  10)]{cover}.  In addition, we must also pay attention to the channel use energetics since lack of
energy constraints can lead to unrealistic results.  Thus, we have defined a channel use as the
launch and capture of $M$ tokens under an emission deadline constraint, $\tau$, with the further
constraint that 
\bee
\label{eq:rhodef}
\mylambda \tau 
= M
\eee
where $\mylambda$, the token launch average intensity, has units of tokens per time.  \Equat{rhodef} is
implicitly a constraint on average power assuming a fixed per-token energy cost for
construction/sequestration/release/delivery.  We also note that the signaling interval $\tau$ is now
an explicit function of $M$ as in
\bd
\label{eq:rhodef2}
\tau = \tau(M) = \frac{M}{\mylambda}
\ed

\begin{figure} [h]
\begin{center}
\includegraphics[height=0.75in,width=3.0in]{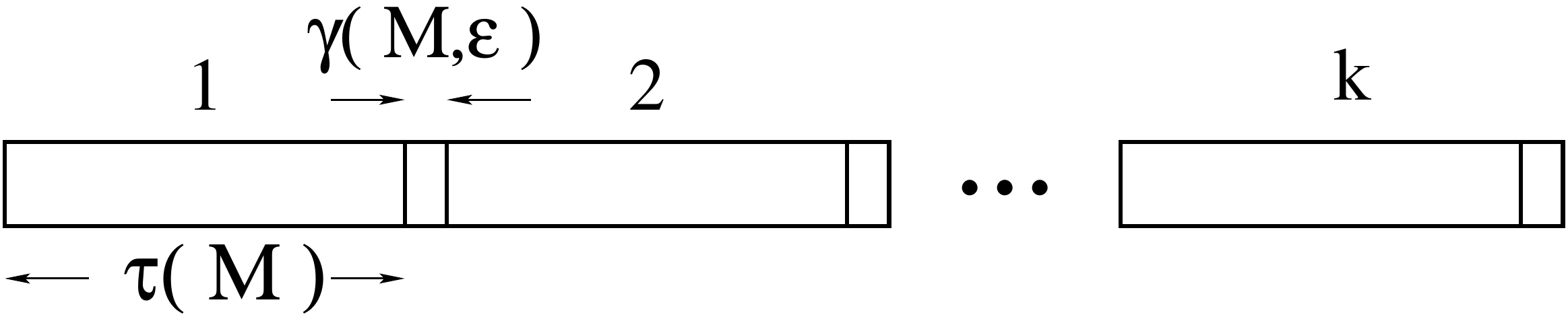}
\end{center}
\vspace{-0.25in}
\caption{Successive $M$-emission channel uses.
For a given use of the token timing channel, the sender emits $M$ tokens over
the transmission interval $\tau(M) = \frac{M}{\mylambda}$.
$\gamma(M,\epsilon)$ is the waiting period (guard interval) before the next channel use.
}
\label{fig:channeluse}
\end{figure}

So, consider FIGURE~\ref{fig:channeluse} where sequential $M$-token transmissions -- channel
uses -- are depicted.  We will assume a ``guard interval'' of some duration $\gamma(M,\epsilon)$
between successive transmissions so that all $M$ tokens are received before the beginning of the
next channel use with probability ($1-\epsilon$) for arbitrarily small $\epsilon > 0$.
We further require that the average emission rate, ${M}/(\tau(M) + \gamma(M,\epsilon))$ satisfies
\bee
\label{eq:limitlambda}
\lim_{\epsilon \rightarrow 0}
\lim_{M\rightarrow \infty}
\frac{M}{\tau(M) + \gamma(M,\epsilon)}
=
\mylambda
\eee
We then require that the last token arrival time $\Svec_M$ occurs before the start of the next channel use with
probability 1.
That is, given arbitrarily small $\epsilon$
we can always find a finite $M^*$ such that
\bee
\label{eq:timelimit1}
\mbox{Prob}\{\vec{S}_M \le \tau(M) + \gamma(M,\epsilon) \} > 1-\epsilon
\eee
$\forall M  \ge  M^*$.
We now derive a sufficient condition on first-passage time densities for which
\equat{timelimit1} is true.

Calculating a CDF for $\vec{S}_M$ is in general difficult since emission times $T_m$ might not be
independent. However, for a fixed emission interval $[0, \tau(M)]$ we can readily calculate a worst
case CDF for $\vec{S}_M$ and thence an upper bound on the guard interval duration that satisfies the
arrival condition of \equat{timelimit1}.  That is, for a given emission schedule $\tv$, the $\Smat$
are conditionally independent and the CDF for the final arrival is
\bee
F_{\Svv_M|\tv}(s|\tv) = \prod_{m=1}^M  G(s-t_m)u(s-t_m)
\eee
so that
\bee
F_{\Svv_M}(s)  = \int_{\bzero}^{{\bf \tau(M)}} f_{\Tmat}(\tv)\prod_{m=1}^M  G (s-t_m)u(s-t_m) d \tv \IEEEeqnarraynumspace
\label{eq:overrun}
\eee
However, it is easy to see that
\bee
F_{\Svv_M}(s) \ge G^M(s-\tau(M)) u(s-\tau(M))
\eee
since $G(s-t)$ is monotone decreasing in $t$.

The end of the guard interval is $\tau(M) + \gamma(M,\epsilon)$, so the probability
that the last arrival time $\Svv_M$ occurs before the next signaling interval obeys
\bee
\label{eq:timelimit2}
F_{\Svv_M}(\tau(M) + \gamma(M,\epsilon))
\ge
G^M \left ( \gamma(M,\epsilon)   \right )
\eee
And to meet the requirement of \equat{timelimit1} we must have
\bee
\label{eq:timelimit2a}
\lim_{M \rightarrow \infty}
G^M \left ( \gamma(M,\epsilon)   \right )
=
1
\eee
which for convenience, we rewrite as
\bee
\label{eq:timelimit3}
\lim_{M \rightarrow \infty}
M \log G \left ( \gamma(M,\epsilon)   \right )
=
0
\eee

If rewrite $\log G \left ( \gamma(M,\epsilon) \right )$ in terms of the CCDF (complementary CDF) $\bG(\cdot)$ (which must be
vanishingly small in large $M$ if we are to meet the conditions of \equat{timelimit1}) and note that $\log(1-x)
\approx -x$ for $x$ small, we have
\bd
- \log \left ( 1 - \bG \left (\gamma(M,\epsilon) \right ) \right )
\approx
\bG \left (\gamma(M,\epsilon) \right )
\ed
for sufficiently large $M$.  Thus, a first-passage distribution whose CCDF satisfies
\bee
\label{eq:timelimit4}
\lim_{M \rightarrow \infty}
M \bG \left (\gamma(M,\epsilon)  \right )
=
0
\eee
with some suitable $\gamma(M,\epsilon)$ will also allow satisfaction of \equat{timelimit1}. However, 
the satisfaction of \equat{timelimit4} requires 
that
$1/\bG \left ( \gamma(M,\epsilon) \right )$ be {\em asymptotically supralinear} in $M$.

We then note that since all first-passage times are non-negative random variables, the mean first-passage
time is given by \cite{papoulis} 
\bee
\label{eq:meanintegral}
E[D]
=
\int_0^{\infty}
\bG(x) dx
\eee
The integral of \equat{meanintegral} exists {\bf iff} $1/\bG(x)$ is asymptotically supralinear in $x$. Thus,
the existence of $E[D]$ in turn implies that
choosing $\gamma(M,\epsilon) = \epsilon M$ allows satisfaction of \equat{timelimit4}.
So, in the limit of vanishing $\epsilon$ we then have
\bd
\lim_{\epsilon \rightarrow 0}
\lim_{M \rightarrow \infty}
\frac{M}{\tau(M) + \gamma(M,\epsilon)}
\ge
\lim_{\epsilon \rightarrow 0}
\frac{\mylambda}{1+\epsilon}
=
\mylambda
\ed
\blankout{
and
\bd
\lim_{\epsilon \rightarrow 0}
\lim_{M \rightarrow \infty}
\frac{M}{\tau(M) + \gamma(M,\epsilon)}
\le
\lim_{\epsilon \rightarrow 0}
\mylambda
= 
\mylambda
\ed
}
and the energy requirement of \equat{limitlambda} is met in the limit while assuring asymptotically
independent sequential channel uses.

The above development proves the following theorem:
{\em   \begin{theorem}{\bf Asymptotically Independent Sequential Channel Uses:}
\thmlabel{channeluse}

Consider the channel use discipline depicted in FIGURE~\ref{fig:channeluse} where tokens are emitted
on an interval $[0,\tau(M)]$ with $\tau(M) = \frac{M}{\mylambda}$ and guard intervals of duration
$\gamma(M,\epsilon)$ are imposed between channel uses.  If the mean first-passage time $E[D]$ is
finite, then guard intervals can always be found such that the sequential channel uses approach asymptotic
independence as $\epsilon \rightarrow 0$, and the relative duration of the guard interval,
$\gamma(M,\epsilon)$ vanishes compared to $\tau(M)$ as $M \rightarrow \infty$.
\end{theorem}}
\begin{Proof}{\Thmref{channeluse}}
See the development leading to the statement of \Thmref{channeluse}.
\end{Proof}

Now, suppose the transport process from source to destination has infinite first passage time,
implying that $1/\bG(x)$ is linear or sublinear in $x$.  Is asymptotically independent sequential
channel use possible? The answer seems to be no.

As a best case, the minimum probability of tokens arriving outside $\tau(M) + \gamma(M,\epsilon)$ is
obtained if all emissions occur at $t=0$ (see \equat{overrun}). Any other token emission
distribution must have larger probability of interval overrun.  For asymptotically independent
sequential channel use we then must have, following \equat{rhodef2} and \equat{timelimit4},
\bee
\label{eq:timelimitsublin}
\lim_{M \rightarrow \infty}
M \bG \left ( \frac{M}{\lambda} + \gamma(M,\epsilon)  \right )
=
0
\eee
We notice that the argument of $\bG(\cdot)$ is at least linear in $M$, and a linear-in-$M$ argument
will not drive $\bG(\cdot)$ to zero faster than $1/M$ because $1/\bG(\cdot)$ is not
supralinear. Thus, the argument of $\bG(\cdot)$ must be supralinear in $M$ to drive $\bG \left (
\frac{M}{\lambda} + \gamma(M,\epsilon) \right )$ to zero faster than $1/M$ which in turn implies
that $\gamma(M,\epsilon)$ must be supralinear in $M$. However, if $\gamma(M,\epsilon)$ is
supralinear in $M$, then \equat{limitlambda} cannot be satisfied and we have proved the
following theorem:
{\em \begin{theorem}{\bf Infinite Mean First Passage Does Not Allow Asymptotically Independent Sequential Channel Uses:}
\thmlabel{infinitemean}

Consider the channel use discipline depicted in FIGURE~\ref{fig:channeluse} where tokens are emitted
on an interval $[0,\tau(M)]$ with $\tau(M) = \frac{M}{\mylambda}$ and guard intervals of duration
$\gamma(M,\epsilon)$ are imposed between channel uses.  If the mean first-passage time $E[D]$ is
infinite, then guard intervals can never be found such that the sequential channel uses approach asymptotic
independence as $\epsilon \rightarrow 0$, and the relative duration of the guard interval,
$\gamma(M,\epsilon)$ vanishes compared to $\tau(M)$ as $M \rightarrow \infty$.
\end{theorem}}
\begin{Proof}{\Thmref{infinitemean}}
See the development leading to the statement of \Thmref{infinitemean}.
\end{Proof}

To summarize, if the mean first passage time exists, then asymptotically independent sequential
channel uses are possible and the mutual information $I(\Svv;\Tmat)$ is the proper measure of
information transport through the channel. Conversely, if the mean first passage time is infinite,
then asymptotically independent sequential channel uses are impossible and the associated channel capacity
problem is ill-posed.  It is worth noting that free-space diffusion (without drift) has infinite
$E[D]$.  However, since all physical systems have finite extent, $E[D]$ is always finite for any
realizable ergodic token transport process.

\subsection{Channel Capacity Definitions}
\label{sect:chandef}
The maximum $I(\Svv;\Tmat)$ is the {\em channel capacity} in units of bits/nats per channel use.
However, we will find it useful to define the maximum mutual information between $\Tmat$ and $\Svv$
{\em per token}.    That is, the channel capacity per token $C_q$ is
\bee
\label{eq:pertokenCtauM}
C_q(M, \tau(M))
\equiv  
\frac{1}{M} 
{\displaystyle \max_{f_{\Tmat}(\cdot)}} I(\Svv;\Tmat)
\eee
Since $\tau(M)= M/\mylambda$, it is easy to see that $C_q(M, \tau(M))$ will be monotone increasing in $M$ since concatenation of two
emission intervals with durations $\tau(M/2)$ and $M/2$ tokens each is more
constrained than a single interval of twice the duration $\tau(M)$ with $M$ tokens.
We can thus say that
\bee
\label{eq:CmGTCmover2}
C_q(M,\tau(M))
\ge
2 C_q(M/2, \tau(M/2))
\eee

We can then define the limiting capacity in nats per token as
\bee
\label{eq:pertokenCtauinf}
C_q
\equiv
\lim_{M \rightarrow \infty}
C_q(M,\tau(M))
\eee
with no stipulation as yet to whether the limit exists or is bounded away from zero.


Now consider the capacity per unit time.  The duration of a channel use (or signaling epoch) is
$\tau(M) + \gamma(M,\epsilon)$ (see FIGURE~\ref{fig:channeluse}).  Thus, for a given number $M$ of
emissions per channel use and a probability $(1-\epsilon)$ that all the tokens are received before
the next channel use, we define the channel capacity in nats per unit time as
\bas
C_t(M,\epsilon) &  \equiv &
{\displaystyle \max_{f_{\Tmat}(\cdot)} \frac{I(\Svv;\Tmat)}{\tau(M) + \gamma(M,\epsilon)}}\\
 & = & 
{\displaystyle C_q(M, \tau(M)) \left (\frac{M}{\tau(M) + \gamma(M,\epsilon)} \right )}
\eas
which in the limits of $\epsilon \rightarrow 0$ and $M \rightarrow \infty$ becomes
\bee
\lim_{\epsilon \rightarrow 0}
\lim_{M \rightarrow \infty}
C_t(M,\epsilon)
=
\lambda C_q
\label{eq:Ctinf}
\eee
via \equat{limitlambda} and \equat{pertokenCtauinf}.

The above development proves the following theorem:
{\em  \begin{theorem}{\bf Capacity of the identical-token timing channel:}
\thmlabel{timelimit_firstpassage}

If the mean first-passage time $E[D]$ exists, then the channel capacity in nats per unit time obeys
\bee
\label{eq:CtislambdaCq}
C_t
=
\mylambda
C_q
\eee
where $C_q$ is the capacity per token defined in \equat{pertokenCtauinf} and $\mylambda$ is the
average token emission rate.
\end{theorem}}
\begin{Proof}{\Thmref{timelimit_firstpassage}}
See \Thmref{channeluse} and the development leading to the statement of \Thmref{timelimit_firstpassage}.
\end{Proof}
It is worth noting that Theorem \thmref{timelimit_firstpassage} is {\em general} and applies to {\em
  any} system with finite first-passage time.  

Now, we more carefully examine the mutual information $I(\Svv;\Tmat)$ to determine whether the
limits implied of \equat{pertokenCtauinf} and \equat{Ctinf} exist and are bounded away from zero.

\subsection{Mutual Information Between Input $\Tmat$ and Output $\Svv$}
\label{sect:mutualinfo}
The mutual information between $\Tmat$ and $\Smat$ is
\bee
\label{eq:unorderedMI}
I(\Smat; \Tmat)
=
h(\Smat) - h(\Smat|\Tmat)
=
h(\Smat) - M h(S|T)
\eee
Since the $S_i$ given the $T_i$ are mutually independent each with density $g(s_i - t_i)$,
$h(\Smat|\Tmat)$ does not depend on $f_{\Tmat}(\tv)$.  Thus, maximization of \equat{unorderedMI} is
simply a maximization of $h(\Smat)$ which is in turn maximized by maximizing the marginal $h(S)$
over the marginal $f_T(t)$, a problem explicitly considered and solved in closed form for a mean
$T_m$ constraint by Anantharam and Verd\'u in \cite{bits-Qs} and for a deadline constraint in \cite{isit11,RoseMian16_2}, both for exponential first-passage.

The corresponding expression for the mutual information between $\Tmat$ and $\Svv$ is
\bee
\label{eq:orderedMI}
I(\Svv; \Tmat)
=
h(\Svv) - h(\Svv|\Tmat)
\eee
Unfortunately, $h(\Svv|\Tmat)$ now {\em does} depend on the input distribution and the maximization
of $h(\Svv)$ is non-obvious.  So, rather than attempting a brute force optimization of
\equat{orderedMI} by deriving order distributions $f_{\Svv}(\cdot)$ \cite{eckford1}, we explore -- {\em
  with no loss of generality} -- simplifying symmetries.

Consider that an emission vector $\tv$ and any of its permutations $P_n(\tv)$ produce statistically
identical outputs $\Svv$ owing to the reordering operation as depicted in
FIGURES~\ref{figure:imchannel} and ~\ref{figure:channel}.  Thus, any $f_{\Tmat}(\cdot)$ which optimizes
\equat{orderedMI} can be ``balanced'' to form an optimizing input distribution which obeys
\bee
\label{eq:Tsymm}
f_{\Tmat} (\tv)
=
f_{\Tmat} (P_n(\tv))
\eee
for $n=1,2,\cdots,M!$ and $P_n(\cdot)$ the previously defined permutation operator (see
\equat{permutationdef}).  We can therefore restrict our search to hypersymmetric densities
$f_{\Tmat}(\tv)$ as defined by \equat{Tsymm}.

Now, hypersymmetric $\Tmat$ implies hypersymmetric $\Smat$ which further implies that
$f_{\Smat}(\sv) = f_{\Smat}(P_k(\sv))$.  The same non-zero corner and folding argument used in the derivation
of \equat{svv_def} produces the following key theorem:
{\em   \begin{theorem}{\bf The entropy $h(\Svv)$ relative to the entropy $h(\Smat)$:}
\thmlabel{orderedS}

If $f_{\Tmat}(\cdot)$ is a hypersymmetric probability density function on emission
times $\{T_m\}$, $m=1,2,..,M$, and the first-passage density $g(\cdot)$ is non-singular,
then the entropy of the time-ordered outputs $\Svv$ is
\bd
h(\Svv) = h(\Smat) - \log M!
\ed
\end{theorem}}
\begin{Proof}{\Thmref{orderedS}}
The hypersymmetry of $f_{\Smat}(\sv)$ implies
\bea
h(\Svv) & =  & -\int_{\svv} M! f_{\Smat}(\svv) \log \left (M! f_{\Smat}(\svv) \right ) d\svv \nonumber\\
 & = & -log M! -\int_{\svv} M! f_{\Smat}(\svv) \log  f_{\Smat}(\svv) d\svv \nonumber\\ 
 & = & -log M! -\int_{\sv}  f_{\Smat}(\sv) \log  f_{\Smat}(\sv) d\sv \nonumber\\ 
 & = & -log M! + h(\Smat)
\eea
\end{Proof}

It is worth noting that hypersymmetric densities  on $\Tmat$ are completely equivalent (from a mutual
information maximization standpoint) to their ``unbalanced'' cousins.  Remember that each and every
$I(\Svv;\Tmat)$-maximizing $f_{\Tmat}(\cdot)$ can be ``balanced'' and made into a hypersymmetric density
without affecting the resulting value of $I(\Svv;\Tmat)$.  Likewise, any hypersymmetric density has
a corresponding ordered density that produces the same $I(\Svv;\Tmat)$.  So, the assumption of
hypersymmetric input densities is simply an analytic aid.

Next we turn to $h(\Svv|\Tmat)$. A zero-measure edge-folding argument on the conditional density is
not easily applicable here, so we resort to some information-theoretic sleight of hand.  As before
we define $\Omega$ as the permutation index number that produces an ordered output from $\Smat$
so that $P_{\Omega}(\Smat) = \Svv$.  We first note the equivalence
\bee
\label{eq:equivalence}
\{ \Omega,\Svv \} \Leftrightarrow \Smat
\eee
That is, specification of $\{ \Omega,\Svv \}$ specifies $\Smat$ and {\em vice versa} because as in
our derivation of $h(\Svv)$, this equivalence requires that we exclude the zero-measure ``edges''
and ``corners'' of the density where two or more of the $s_i$ are equal. Thus, there is no
ambiguity in the $\Smat \rightarrow \Svv$ map.

We then have,
\bee
\label{eq:jointdiscretedef}
h(\Smat|\Tmat)
=
h(\Omega, \Svv|\Tmat)
=
h(\Svv|\Tmat)
+
H(\Omega|\Svv, \Tmat)
\eee
which also serves as an {\em en passant} definition for the entropy of a joint mixed distribution
($\Omega$ is discrete while $\Sv$ is continuous).  We then rearrange \equat{jointdiscretedef} to
prove a key theorem:
{\em   \begin{theorem}{\bf The Ordering Entropy, $H(\Omega|\Svv,\Tmat)$:}
\thmlabel{Homega}

\bee
\label{eq:Homega}
h(\Svv|\Tmat)
=
h(\Smat|\Tmat)
-
H(\Omega|\Svv, \Tmat)
\eee
where $H(\Omega|\Svv, \Tmat)$, the {\em ordering entropy},  is the uncertainty about which $S_m$ corresponds to which $\vec{S}_m$ given both $\Tmat$ and $\Svv$.
\end{theorem}}
\begin{Proof}{\Thmref{Homega}}
See \equat{jointdiscretedef}.
\end{Proof}
We note that 
\bee
\label{eq:HOeasybounds}
0 \le H(\Omega|\Svv, \Tmat) \le \log M!
\eee
with equality on the right for any singular density, $f_{\Tmat}(\cdot)$, where all the $T_m$ are equal
with probability $1$.  We can then, after assuming that $f_{\Tmat}(\cdot)$ is hypersymmetric, write the
ordered mutual information in an intuitively pleasing form:
{\em  \begin{theorem}{\bf The mutual information $I(\Svv;\Tmat)$ relative to the
mutual information $I(\Smat;\Tmat)$:}
\thmlabel{Isvv}

For a hypersymmetric density $f_{\Tmat}(\tv) = f_{\Tmat}(P_k(\tv))$, $k=1,2, \cdots, M!$, the mutual information between launch times $\Tmat$ and ordered arrival times $\Svv$ satisfies
\bee
\label{eq:orderedMI_decomp}
I(\Svv; \Tmat)
=
I(\Smat;\Tmat)
-
\left ( \log M! - H(\Omega|\Svv,\Tmat) \right )
\eee
\end{theorem}}
\begin{Proof}{\Thmref{Isvv}}
Combine \Thmref{orderedS} and \Thmref{Homega} with \equat{orderedMI}.
\end{Proof}

Put another way, an average {\em information degradation} of $\log M! - H(\Omega|\Svv,\Tmat) \ge 0$
is introduced by the sorting operation, $\Smat \rightarrow \Svv$.

Mutual information is convex in $f_{\Tmat}(\tv)$ and the space ${\cal
  F}_{\Tmat}$ of feasible hypersymmetric $f_{\Tmat}(\tv)$ is convex. That is,
for any two hypersymmetric probability functions $f_{\Tmat}^{(1)}$ and
$f_{\Tmat}^{(2)}$ we have
\bee
\label{eq:convexf_T}
\kappa f_{\Tmat}^{(1)}(\tv) + (1-\kappa) f_{\Tmat}^{(2)}(\tv)
\in {\cal F}_{\Tmat}
\eee
where $0 \le \kappa \le 1$.  Thus, we can in principle apply variational
\cite{hild} techniques to find that hypersymmetric $f_{\Tmat}(\cdot)$ which attains
the unique maximum of \equat{orderedMI_decomp}. However, in practice, direct
application of this method leads to grossly infeasible $f_{\Tmat}(\cdot)$,
implying that the optimizing $f_{\Tmat}(\cdot)$ lies along some ``edges'' or in some
``corners'' of the convex search space.

\subsection{An Analytic Bound for Ordering Entropy $H(\Omega|\Svv, \Tmat)$}
The maximization of \equat{orderedMI_decomp} hinges on specification of $H(\Omega|\Svv,\Tmat)$, the
{\em ordering entropy} given $\Svv$ and $\Tmat$.  To determine analytic expressions for
$H(\Omega|\Svv,\Tmat)$, consider that given $\tv$ and $\svv$, the probability that $\svv$ was
produced by the $k^{\mbox{th}}$ permutation of the underlying $\sv$ is
\bee
\label{eq:Pst}
\mbox{Prob}(\Omega = k| \svv, \tv)
=
\frac{f_{\Smat|\Tmat}(P_k^{-1}(\svv)| \tv)}
{\displaystyle{\sum_{n=1}^{M!}}
f_{\Smat|\Tmat}(P_n(\svv)| \tv)}
\eee
where $\svv = P_{k}(\sv)$.  Some permutations will have zero probability (are {\em inadmissible})
since the specific $\svv$ and $\tv$ may render them impossible via the causality of $g(\cdot)$.

Using \equat{svvT_def}, the definition of entropy, and \equat{Pst} we have
\bea
\IEEEeqnarraymulticol{3}{l}{H(\Omega|\svv,\tv)}   \nonumber  \\ \quad
 & = & - \!\!\sum_{n=1}^{M!}
\left  [
\frac{\gv(P_n(\svv) \!-\! \tv)}{\displaystyle{\sum_{j=1}^{M!}}\gv(P_j(\svv) \!- \!\tv)}
\right ]
\!\log
\!\!\left [
{\frac{\gv(P_n(\svv) \!- \!\tv)}{\displaystyle{\sum_{j=1}^{M!}}\gv(P_j(\svv)\! -\! \tv)}}
\right ] \IEEEeqnarraynumspace
\label{eq:HstG}
\eea
and as might be imagined, \equat{HstG} is difficult to work with in general.  Nonetheless, let us
define the number of nonzero terms in the sum of \equat{HstG} as $|\Omega|_{\svv,\tv}$.

Now, consider that for exponential $g(\cdot)$, we can use \equat{econd} to write \equat{Pst} as
\bee
\label{eq:PstE}
\mbox{Prob}(\Omega = k| \svv, \tv)
=
\frac{\uv(P_k^{-1}(\svv)- \tv)}
{\displaystyle{\sum_{n=1}^{M!}}
\uv(P_n(\svv)- \tv)}
\eee
where $\uv(\cdot)$ is a multidimensional unit step function.  \Equat{PstE} is a {\em uniform} probability
mass function with $\sum_{n=1}^{M!}\uv(P_n(\svv)- \tv) = |\Omega|_{\svv,\tv}$ elements -- the same as the number of
non-zero terms in the sum of \equat{HstG}.  Thus, 
\bee
\label{eq:He}
H(\Omega|\svv,\tv)
\le
\log 
\displaystyle{\sum_{n=1}^{M!}}
\uv(P_n(\svv)- \tv)
\eee
for {\em all possible causal first-passage time densities}, $g(\cdot)$. In addition, it can be shown
that exponential first-passage time is the {\em only} first-passage density which maximizes
$H(\Omega|\svv,\tv)$, a result we state as a theorem:
{\em   \begin{theorem}{\bf A General Upper Bound for $H(\Omega|\svv,\tv)$:}
\thmlabel{Homegaupperbound}

If we define the number of admissible combinations $\{ P_n(\svv), \tv \}$ as
\bd
\left |
\Omega
\right |_{\svv,\tv}
\equiv
\sum_{n=1}^{M!}
\uv(P_n(\svv)- \tv)
\ed
where $\uv(\cdot)$ is a multidimensional unit step function, 
then
\bd
H(\Omega|\svv,\tv)
\le
\log 
\left |
\Omega
\right |_{\svv,\tv}
\ed
with equality {\bf iff} $g(\cdot)$ is exponential.
\end{theorem}}
\begin{Proof}{\Thmref{Homegaupperbound}}
We have already shown via \equat{PstE} that exponential first passage renders $\mbox{Prob}(\Omega = k|
\svv, \tv)$ uniform.

Now, consider that the probability mass  function (PMF) of \equat{Pst} can be written as
\bd
\mbox{Prob}(\Omega = k| \svv, \tv)
=
\frac{\gv(P_k^{-1}(\svv) - \tv)}{\displaystyle{\sum_{j=1}^{M!}}\gv(P_j(\svv) - \tv)}
\ed
This PMF is uniform {\bf iff} for all $n$ and $k$ where $P_n(\svv)$ and $P_k(\svv)$ {\em are both causal with respect to $\tv$} we have
\bee
\label{eq:permute}
\gv(P_n(\svv) - \tv)
=
\gv(P_k(\svv) - \tv)
\eee
That is, \equat{permute} must hold for all pairs $(P_n(\svv), \tv)$ and $(P_k(\svv), \tv)$ that are
{\em admissible}. Since the maximum number of non-zero probability $\Omega$ is exactly the
cardinality of admissible $(P_n(\svv), \tv)$, any density which produces a uniform PMF over
admissible $\Omega$ thereby maximizes $H(\Omega|\svv,\tv)$, which proves the inequality.

We then note that any given permutation of a list can be achieved by sequential pairwise swapping of
elements.  Thus, \equat{permute} is satisfied {\bf iff}
\bee
\label{eq:product}
g(x_1 - t_1)g(x_2-t_2)
=
g(x_2 - t_1)g(x_1-t_2)
\eee
$\forall$ admissible $\{ (x_1,x_2)$,  $(t_1,t_2)  \}$.  Rearranging \equat{product} we have
\bd
\frac{g(x_1-t_1)}{g(x_1-t_2)}
=
\frac{g(x_2-t_1)}{g(x_2-t_2)}
\ed
which implies that
\bd
\frac{g(x-t_1)}{g(x-t_2)}
=
\mbox{Constant w.r.t. $x$}
\ed

Differentiation with respect to $x$ yields
\bd
\frac{g^\prime(x-t_1)}{g(x-t_2)}
-
\frac{g(x-t_1)g^\prime(x-t_2)}{g^2(x-t_2)}
=0
\ed
which we rearrange to obtain
\bd
\frac{g^\prime(x-t_1)}{g(x-t_1)}
=
\frac{g^\prime(x-t_2)}{g(x-t_2)}
\ed
which further implies that
\bee
\label{eq:Ediffeq}
\frac{g^\prime(x-t_1)}{g(x-t_1)}
=
c
\eee
since $t_1$ and $t_2$ are free variables.  The only solution to
\equat{Ediffeq} is
\bd
g(x) \propto e^{cx}
\ed
Thus, exponential $g(\cdot)$ is the only first-passage time density that can produce a
maximum cardinality uniform distribution over $\Omega$ given $\svv$ and $\tv$ --
which completes the proof.
\end{Proof}

Now consider that $\left | \Omega \right |_{\svv,\tv}$, as defined in \Thmref{Homegaupperbound}, is
a hypersymmetric function of $\svv$ and $\tv$ and thus invariant under any permutation of its
arguments $\svv$ or $\tv$. That is,
\bas
{\sum_{n=1}^{M!}}
\uv(P_n(\svv) - \tv)
 & = &
{\displaystyle \sum_{n=1}^{M!}}
\uv(P_n(\svv)- \tvv) \\
 & = & 
{\displaystyle \sum_{n=1}^{M!}}
\uv(P_{n}(\sv)- \tvv) \\
 & = & 
{\displaystyle \sum_{n=1}^{M!}}
\uv(P_n(\sv)- \tv)
\eas
because the summation is over all $M!$ permutations.  Therefore, 
\bee
\label{eq:symmetry}
\left |
\Omega
\right |_{\svv,\tv}
=
\left |
\Omega
\right |_{\svv,\tvv}
=
\left |
\Omega
\right |_{\sv,\tvv}
=
\left |
\Omega
\right |_{\sv,\tv}
\eee

We must now enumerate this number of admissible permutations.  Owing to \equat{symmetry} and
\Thmref{Ehypersymmetry} we can assume time-ordered inputs $\tvv$ with no loss of generality. So, let
us define contiguous ``bins'' ${\cal B}_k = \{t| t \in [\tvec_k, \tvec_{k+1}) \}$, $k=1,2,...,M$
  ($\tvec_{M+1} \equiv \infty$) and then define $\sigma_m$ as bin occupancies.  That is, $\sigma_m = q$
  if there are exactly $q$ arrivals in ${\cal B}_m$.  The benefit of this approach is that the
  $\sigma_m$ do not depend on whether $\svv$ or $\sv$ is used to count the arrivals.  Thus,
  expectations can be taken over $\Smat$ whose components are mutually independent given the $\tv$
  and no order distributions for $\Svv$ need be derived.

To determine the random variable $\left | \Omega \right |_{\Smat,\tvv}$ we start by defining
\bd
\eta_m = \sum_{j=1}^{m} \sigma_j
\ed
the total number of arrivals up to and including bin ${\cal B}_m$.
Clearly $\eta_m$ is monotonically increasing in $m$ with $\eta_0 = 0$ and
$\eta_M = M$.  We then observe that the $\sigma_m$ arrivals on $[\tvec_m,\tvec_{m+1})$ can be
assigned to any of the $\tvec_1, \tvec_2, ..., \tvec_m$ known emission times {\em except}
for those $\eta_{m-1}$ previously assigned.  The number of possible new assignments
is $(m-\eta_{m-1})!/(m - \eta_m)!$ which when applied iteratively leads to
\bee
\label{eq:Omegacardb}
\left |
\Omega
\right |_{\Smat,\tvv}
=
\prod_{m=1}^M
\frac{(m - \eta_{m-1})!}{(m - \eta_{m})!}
=
\prod_{m=1}^{M-1}
(m + 1 - \eta_m)
\eee

We then define the random variable
\bd
X_i^{(m)}
=
\twodef{1}{S_i < \tvec_{m+1}}{0}{\mbox{otherwise}}
\ed
for $i=1,2,...m$.  The PMF of $X_i^{(m)}$ is then
\bee
p_{X_i^{(m)}}(x)
=
\twodef{G(\tvec_{m+1} - \tvec_i)}{x = 1}{\bar{G}(\tvec_{m+1} - \tvec_i)}{x = 0}
\label{eq:probX}
\eee
where we note that for a given $m$, $X_i^{(m)}$ and $X_j^{(m)}$ are independent, $i \ne j$, and as
previously defined, $G(\cdot)$ is the CDF of the causal first-passage density $g(\cdot)$.
$\bar{G}(\cdot) = 1 - G(\cdot)$ is the corresponding CCDF.  We can then write
\bd
\eta_m = \sum_{i=1}^m  X_i^{(m)}
\ed

It is then convenient to define $\bar{X}_i = 1 - X_i$ which allows us to define $\bar{\eta}_m = m -
\eta_m$.  We can then write
\bee
\label{eq:Omegacardbbar}
\left |
\Omega
\right |_{\Smat,\tvv}
=
\prod_{m=1}^{M-1}
(1 + \bar{\eta}_m)
\eee
Since we seek the expected value of \equat{Omegacardbbar}, we can use
\equat{probX} to calculate each individual $E_{\Smat|\tvv} [\log (1 + \bar{\eta}_m) ]$ as
\bee
\label{eq:Homega_etabar}
\sum_{\bar{\xv}}
\log (1 \! + \! \sum_{i=1}^m \! \bar{x}_i)
\prod_{j=1}^m
\bar{G}^{\bar{x}_j}({\tvec_{m+1} \! - \! \tvec_j})
G^{1 -\bar{x}_j}({\tvec_{m+1} \! - \! \tvec_j}) \IEEEeqnarraynumspace
\eee
which allows us to define $\Hup(\tv)$, an upper bound on $H(\Omega|\Svv,\tv)$, as
\bea
{\Hup(\tv)}  & \equiv & \sum_{m=1}^{M-1} \sum_{\bar{\xv}} \log (1 + \sum_{i=1}^m \bar{x}_i) \nonumber \\
 & & \times \>{\prod_{j=1}^m \bar{G}^{\bar{x}_j}(\tvec_{m+1} - \tvec_j)
G^{1 -\bar{x}_j}(\tvec_{m+1} - \tvec_j)}
\label{eq:Hexpurg}
\eea
where an ordering permutation on $\tv$ is part of the function $\Hup(\tv)$.
\Equat{Hexpurg} can be rearranged as
\bea
\IEEEeqnarraymulticol{3}{l}{\Hup(\tv) = \sum_{\ell=1}^{M-1} \log(1 + \ell)} \nonumber \\ \quad
& \times & 
{\sum_{m=\ell}^{M-1}}
{ \sum_{|\bar{\xv}| = \ell}} { \prod_{j=1}^m 
\bG^{\bar{x}_j}(\tvec_{m+1} - \tvec_j)
G^{1 -\bar{x}_j}(\tvec_{m+1} - \tvec_j)}  \IEEEeqnarraynumspace
\label{eq:Hexpurg3}
\eea

We then note that
\bd
\begin{array}{rcl}
H(\Omega|\Svv,\Tmat) & \equiv  & E_{\Tmat} \left [ E_{\Svv|\Tmat}  [ H(\Omega|\svv, \tv)  ]  \right ]\\
   & \le   & 
E_{\Tmat} \left [E_{\Svv|\Tvv}  [ \log \left | \Omega  \right |_{\svv,\tv} ]  \right ]\\
 & = & 
E_{\Tvv} \left [E_{\Smat|\Tvv}  [ \log \left | \Omega  \right |_{\sv,\tvv}  ] \right ]
\end{array}
\ed
follow from \equat{Hexpurg3} in conjunction with \Thmref{Homegaupperbound} and through
hypersymmetric expectations  (\Thmref{Ehypersymmetry}) of hypersymmetric functions $\left | \Omega \right |_{\sv,\tvv}$ (\equat{symmetry}).   Adding in the result of \Thmref{Homegaupperbound}
we have proven the following theorem:
{\em  \begin{theorem} {\bf A General and Computable Upper Bound for $H(\Omega|\Svv,\Tmat)$: }
\thmlabel{Hbound}

\bee
E \left [ H(\Omega|\svv,\tv) \right ]
\equiv
H(\Omega|\Svv,\Tmat)
\le
\Hup(\Tmat)
\eee
with equality {\bf iff} the first-passage time density  $g(\cdot)$ is exponential.
\end{theorem}}
\begin{Proof}{\Thmref{Hbound}}
See the development leading to the statement of \Thmref{Hbound}. \Thmref{Homegaupperbound} establishes equality {\bf iff} the first passage density is exponential.
\end{Proof}

Theorem~\ref{thm:Hbound} gives us $\Hup(\Tmat)$, a computable analytic upper bound for $H(\Omega|\Svv,\Tmat)$, and
an exact expression if the first-passage time is exponential.

\subsection{Capacity Bounds For Timing Channels}
\label{sect:timingcap}
Despite significant effort, direct optimization of mutual information, $I(\Svv;\Tmat)$ (see
\equat{orderedMI_decomp}) remained elusive. The key issue is that $h(\Smat)$ and
$H(\Omega|\Tmat,\Svv)$ are ``conflicting'' quantities with respect to $f_{\Tmat}(\cdot)$.  That is,
independence of the $\{ T_m \}$ favors larger $h(\Smat)$ (i.e., $h(\Smat) \le \sum_m h(S_m)$) while
tight correlation of the $\{ T_m\}$ (as in $T_i=T_j$, $i,j = 1,2,...,M$) produces the maximum
$H(\Omega|\Svv,\Tmat) = \log M!$.  In light of these difficulties, we sought analytic expressions
in the companion to this paper (Part~II  {\cite{RoseMian16_2}}) for $h(\Smat)$ and $I(\Smat;\Tmat)$ 
which we restate here as \Thmref{maxhs} and \Thmref{minmaxexp} without proof.

{\em \begin{theorem}{\bf Maximum $h(s)$ for exponential first passage under a deadline constraint:}
\thmlabel{maxhs}

For first-passage time $D$ with density $f_D(d)= g(d) = \mymu e^{-\mymu d}$, and 
launch time $T$ constrained to $[0,\tau]$, the maximum entropy of $S=T+D$ is
\bee
\label{eq:minmaxhS}
\max_{f_T(\cdot)} h(S)
=
\log \left ( \frac{e + \mymu \tau}{\mymu} \right )
\eee
The input density $f_T(\cdot)$ which produces the maximum $h(S)$ is
\bee
\label{eq:expoptfT}
\begin{array}{rcl}
f_T(t) & = &
\delta(t) \frac{1}{e + \mymu \tau} + \delta(t-\tau) \frac{1-e}{e+\mymu \tau}\\
 & + & \frac{\mymu}{e+\mymu \tau} (u(t) - u(t-\tau))
\end{array}
\eee
\end{theorem}}

{\em   \begin{theorem}{\bf Maximum $I(S;T)$ For Exponential First Passage Under A Deadline Constraint:}
\thmlabel{minmaxexp}

For first-passage time $D$ with density $f_D(d)= g(d) = \mymu ^{-\mymu d}$, and 
launch time $T$ constrained to $[0,\tau]$, the maximum mutual information between $S=T+D$ and $T$ is
\bee
\label{eq:inequality}
\max_{f_T}
I(S;T) 
=
\log \left (1 +  \frac{\mymu \tau}{e} \right )
\eee
\end{theorem}}

The definition of $C_q$ and $C_t$ in \Thmref{timelimit_firstpassage} requires we consider the
asymptotic value of $H(\Omega|\Svv,\Tmat)/M$.  A lower bound is provided in Part-II
\cite{RoseMian16_2} assuming exponential first passage, a result we restate here as
\Thmref{HOasympt} without proof.

{\em  \begin{theorem} {\bf \boldmath Asymptotic $H(\Omega|\Svv,\Tmat)/M$ For Exponential First-Passage Under A Deadline Constraint {\boldmath $\Tmat \in [\bzero,\btau]$}:}
\thmlabel{HOasympt}

For exponential first-passage with mean $1/\mymu$, token launch intensity $\mylambda$, and i.i.d. input
distribution $f_{\Tmat}(\tv) = \prod_{m=1}^M f_T(t)$ where $f_T(\cdot)$ maximizes $I(S;T)$ as in
\Thmref{minmaxexp}, the asymptotic ordering entropy per token is
\bee
\label{eq:limHO}
\lim_{M \rightarrow \infty}
\frac{H(\Omega|\Svv,\Tmat)}{M}
=
\sum_{k=2}^{\infty}
e^{-\myrho}
\frac{\myrho^k}{k!}
(\frac{k}{\myrho} - 1)
\log k!
\eee
where $\myrho = \mylambda/\mymu$ is defined as a measure of system token ``load'' similar to a
queueing system.
\blankout{\Equat{limHO} can also be rewritten as an expectation over a Poisson random variable $K$ with parameter $\myrho$:
\bee
\label{eq:limHOPoisson}
\sum_{k=0}^{\infty}
e^{-\myrho}
\frac{\myrho^k}{k!}
(\frac{k}{\myrho} - 1)
\log k!
=
E_K \left [
  (\frac{K}{\myrho} - 1)
  \log K! \right ]
\eee}
\end{theorem}}

We can rewrite the summation term in \equat{limHO} more compactly noting that
\bee
{\displaystyle \sum_{k=1}^{\infty}} \frac{\myrho^k}{k!}
(\frac{k}{\myrho} - 1)
\log k!  = {\displaystyle \sum_{\ell=1}^{\infty}}
\log \ell
{\displaystyle \sum_{k=\ell}^{\infty}}
\frac{\myrho^k}{k!}
{(\frac{k}{\myrho} - 1)}
\eee
Then
\bee
\sum_{k=\ell}^{\infty}
\myrho^k \frac{1}{k!}
=
e^{\myrho}
-
\sum_{k=0}^{\ell - 1}
{\myrho}^k \frac{1}{k!}
\eee
and
\bd
\sum_{k=\ell}^{\infty} 
 \frac{k}{\myrho} \myrho^k \frac{1}{k!}
=
\sum_{k=\ell-1}^{\infty} 
\myrho^k \frac{1}{k!}
\ed
\blankout{
which leads to
\bd
\sum_{k=\ell}^{\infty}
(\frac{k}{\myrho} - 1)\left  \myrho^k \frac{1}{k!} 
=
\sum_{k=\ell-1}^{\infty} 
\myrho^k \frac{1}{k!}
+
\sum_{k=0}^{\ell - 1}
\myrho^k \frac{1}{k!}
-
e^{\myrho}
\ed}
can be used to obtain
\bd
\sum_{k=\ell}^{\infty}
\frac{\myrho^k}{k!}  {(\frac{k}{\myrho} - 1)}
=
\frac{1}{(\ell - 1)!} \myrho^{\ell -1}  
=
\ell \myrho^{\ell}  \frac{1}{\ell!\myrho} 
\ed
We then note that 
\bee
\label{eq:p_ell}
p_\ell
=
e^{-\myrho}
\frac{\myrho^\ell}{\ell!}
\eee
$\ell = 0, 1, \cdots, \infty$ is a Poisson probability mass function and obtain the more compact
\bee
\label{eq:HOsimple}
{\displaystyle \sum_{k=1}^{\infty}}
\myrho^k
(k/\myrho - 1)
\frac{\log k!}{k!}
=
\frac{1}{\myrho} E_\ell \left [ \ell \log \ell \right ]
\eee

Now turning toward capacity, \equat{orderedMI_decomp} and \Thmref{minmaxexp} are easily combined to show
\bd
\frac{1}{M} I(\Smat;\Tmat) - \frac{1}{M} \log M!
\ge
\log \left (1 +  \frac{\mymu \tau}{e} \right ) - \frac{1}{M}\log M!
\ed
Then, since $\tau = M/\mylambda$ we have
\bee
\label{eq:left}
\lim_{M \rightarrow \infty}
\frac{1}{M} I(\Smat;\Tmat) - \frac{1}{M} \log M!
\ge
\log \frac{1}{\myrho}
\eee
Noting that $I(\Svv;\Tmat) = I(\Smat;\Tmat) - \log M! + H(\Omega|\Svv,\Tmat)$ and
$H(\Omega|\Svv,\Tmat)\ \ge 0$ proves the following theorem:
{\em \begin{theorem} {\bf \boldmath A Simple Lower Bound for $C_q$ under exponential first passage:}
  \thmlabel{IDlowerbound}

Given a token launch intensity $\mylambda = M/\tau$ and exponential first-passage time distribution
with mean $\mymu^{-1}$, the timing channel capacity $C_q(\myrho)$ in nats per token obeys
\bee
\label{eq:Cqlo}
C_{q}(\myrho) \ge \max \left
    \{-\log\myrho,0 \right \}
\eee
where $\myrho = \frac{\mylambda}{\mymu}$
\end{theorem}}
\begin{Proof}{\Thmref{IDlowerbound}}
See the development leading to the statement of \Thmref{IDlowerbound}.
\end{Proof}

We can, however, combine \equat{left}, \Thmref{HOasympt} and \equat{HOsimple} to obtain a better lower
bound on capacity:
{\em \begin{theorem}{\bf Lower Bound for $C_q$ and $C_t$ for Exponential First Passage:}
\thmlabel{CqCtloexp}

In the limit of large $M$, with mean $1/\mymu$ exponential first-passage, the  channel capacities, $C_q$ and
$C_t$ must obey
\bee
C_q(\myrho)
\ge
\log \frac{1}{\myrho}
+
\frac{1}{\myrho} E \left [ \ell  \log \ell \right ]
\eee
and
\bee
C_t(\myrho)
\ge 
\mylambda \left (
\log \frac{1}{\myrho}
+
\frac{1}{\myrho} E \left [ \ell  \log \ell \right ] \right )
\eee
where $\ell$ is Poisson with PMF
\bee
p_{\ell}
=
e^{-\myrho}
\frac{\myrho^\ell}{\ell!}
\eee
where $\myrho = \mylambda/\mymu$.
\end{theorem}}
\begin{Proof}{\Thmref{CqCtloexp}}
Combine \equat{left}, \Thmref{HOasympt} and \equat{HOsimple}.
\end{Proof}

Finally, from Part-II \cite{RoseMian16_2, isit14} we have the following upper bound on $C_q$ (and the concomitant bound on $C_t = \mylambda C_q$) as:
{\em \begin{theorem}{\bf Upper Bound for $C_q$ and $C_t$ for Exponential First Passage:}
\thmlabel{Iupbound}

If the first-passage density $f_D(\cdot)$ is exponential with parameter $\mymu$ and the
rate at which tokens are released is $\mylambda$, then the capacity per
token, $C_q$ is upper bounded by
\bee
\label{eq:Cmup}
C_q
\le
\log \left (\frac{1}{\myrho} +  4 \right )
\eee
and the capacity per unit time is upper bounded by 
\bee
\label{eq:Ctup}
C_t
\le
\mylambda
\log \left (\frac{1}{\myrho} +  4 \right )
\eee
where $\myrho =  \frac{\mylambda}{\mymu}$
\end{theorem}}
\begin{Proof}{\Thmref{Iupbound}}
 Theorem 11 in Part-II \cite{RoseMian16_2} provides the bound for $C_q$ and application of
\Thmref{timelimit_firstpassage} provides the bound for $C_t$.
\end{Proof}

We have now concluded our treatment of the identical token timing channel.  In the next sections we
show how these results can be applied to molecular communication channels where tokens can carry
information payloads.

\subsection{Tokens With Payloads}
\label{sect:payload}
In sections~\ref{sect:channeluse} through \ref{sect:timingcap} we developed all the machinery
necessary to provide capacity bounds for channels with identical tokens where timing is the only
means of information carriage. However, one can also imagine scenarios where the token itself
carries information, much as a ``packet'' carries information over the Internet.  That is, assume
the token is a finite string of symbols over a finite alphabet.  Having constructed tokens from
these ``building blocks,'' a sender launches them into the channel and they are captured by a
receiver. In this scenario a DNA sequence is a symbolic string drawn from a 4-character alphabet so
that each nucleotide could carry 2 bits of information. Similarly, a protein sequence is a symbolic
string drawn from a 20-character alphabet so that each amino acid could carry a little over 4 bits
of information. Thus, a DNA token constructed from $100$ nucleotides would carry $200$ bits whereas
a corresponding protein token would carry $>400$ bits.

However, there are myriad other possibilities for coding information in structure.  For example, a
third major class of biological macromolecules, carbohydrates (polysaccharides), are linear and
branched polymers constructed from a larger alphabet of monosaccharides.  In addition to the
composition information inherent in the makeup of a linear or non-linear concatenations of building
blocks, one could imagine a layer of structural information as well \cite{Bro03} -- as is the case
with biological macromolecules where the spatiotemporal architecture of a polymer is as important as
the order and frequencies of nucleotide, amino acid or monosaccharide residues in the sequence
string.  However, as the issue of ``structural'' information (the amount of information contained in
an arbitrary 3-dimensional object) is as yet an open problem, we will not consider such
constructions in detail.  Nonetheless, the bounds we will derive are applicable to {\em any} method
of information transfer wherein tokens carry information payload, either as string sequences or in
some other structural way.

So, for now consider only string tokens -- as exemplified by DNA and protein sequences -- where each
token in the ensemble released by the sender carries a portion of the message. Thus, irrespective of
their individual lengths, such ``inscribed matter'' tokens must be ``strung together'' to recover
the original message, which implies that {\em each token must be identifiable}. Just as in human
engineered systems like the Internet where information packets could arrive out of order, a sequence
number could be appended to each packet to ensure proper reconstruction at the destination.  Thus,
given $M$ tokens per channel use, we could append $\log M$ bits to each token. We will defer
detailed discussion of this scenario until section~\ref{sect:discussion} as this approach is
asymptotically impractical with $\log M$ tending toward $\infty$.  Alternatively, one could employ
gross differences to convey sequence information such as sending tokens of distinct lengths
$1,2,\cdots, K$ where $M= K(K+1)/2$, and there may be other clever ways to embed structural
side-information to establish token sequence.  Nonetheless, the myriad possibilities
notwithstanding, $H(\Omega|\Svv,\Tmat)$ provides the measure of essential token ``overhead'' or
``side-information'' (of any form) necessary to maintain proper sequence.

Consider that operation of the timing channel involves construction of ``blocks'' $\{\tv_1, \cdots,
\tv_N \}$ where each $\tv_n$ represents the launch schedule for $M$ tokens (a channel use). These
blocks, ``codewords'' of blocklength $N$, are launched into the channel. If capacity is not
exceeded, the receiver can reliably recover the information embedded in the codewords and since we
generally assume the receiver has access to the coding method, a correctly decoded message implies
knowledge of the codewords $\{\tv_1, \cdots, \tv_N \}$.  However, the channel imposes residual
uncertainty about the mapping $\Smat \rightarrow \Svv$ -- the ambiguity about which $\svec_i$ is
associated with which $s_j$.  For this reason, the payload-inscribed tokens cannot yet be correctly
strung together to recover the message.

However, given the observed arrivals $\svv$ and the correctly decoded $\tv$, $H(\Omega|\svv, \tv)$
is {\em the definition} of the uncertainty about that ordering, $\Omega$.  Likewise, the average
uncertainty is $H(\Omega|\Svv,\Tmat)$.  Thus, the source coding theorem implies
that at least $H(\Omega|\Svv,\Tmat)$ bits must be used, on average, to resolve the mapping
ambiguity.

So, consider a message $\Pmat$ to be carried as token payload that we break into equal size ordered
submessages $p_m$, $m=1,2,\cdots, M$.  We can summarize the previous discussion as a theorem:
{\em  \begin{theorem} {\bf Sequencing Information for Tokens with Payload:}
\thmlabel{Hseq}

If a message $P$ is broken into equal size ``payload'' submessages $\{ p_m \}$, $m=1,2,\cdots, M$ and inscribed
into otherwise identical tokens launched at times $\{t_m\}$, we must provide, on average, additional
``sequencing information'' $\frac{1}{M}H(\Omega|\Svv,\Tmat)$ per token at the receiver to assure
recovery of the full payload message $\Pmat = p_1p_2 \cdots p_M$.
\end{theorem}}
\begin{Proof}{\Thmref{Hseq}}
Given arrivals $\svv$ and known departures $\tv$, the uncertainty about the mapping between the $\{
\svec_m \}$ and the $\{ s_m\}$ (and thus the associated $\{ t_m \}$) is exactly the ordering entropy
$H(\Omega|\svv, \tv)$.  Considering $\Omega$ as a letter from a random i.i.d. source, the source
coding theorem \cite{cover, gallagerit} requires at least $H(\Omega|\svv, \tv)$ bits on average to
uniquely specify $\Omega$ -- or asymptotically over many channel uses, at least
$H(\Omega|\Svv,\Tmat)$ bits on average.  Therefore the information necessary at the receiver to
recover the proper sequence and thence the message $P$ is greater than or equal to
$\frac{1}{M}H(\Omega|\Svv,\Tmat)$.
\end{Proof}

It is important to note that we have not actually provided a {\em method} for message
reconstruction, only a lower bound on the amount of ``side information'' necessary 
at the receiver to assure proper reconstruction.  However, as a practical matter, the quantity
$\frac{1}{M}H(\Omega|\Svv,\Tmat)$ does provide some guidance.  In the worst case where the order of
token arrival is completely random, $H(\Omega|\Svv,\Tmat) = \log M!$ which amounts to each packet
carrying a header of size $\frac{1}{M} \log M! \approx \log M$ for large $M$ -- essentially
numbering the packets from $1$ to $M$.  If $\frac{1}{M}H(\Omega|\Svv,\Tmat)$ is much smaller, then
one could imagine cyclic packet numbers since smaller $\frac{1}{M}H(\Omega|\Svv,\Tmat)$
implies that packets launched far apart in time are unlikely to arrive out of order.  The sequence header
could then be commensurately smaller.  In either case, the total amount information necessary to resolve the ordering $\Omega$ is $\frac{1}{M}H(\Omega|\Svv,\Tmat)$ on average.

\subsection{Energy Costs}
\label{sect:energy}

System energy is a critical resource which limits capacity in all communication systems.  In the
case of molecular communication, there are a variety of potential costs, most notably manufacture,
launch and transport. So, assume the minimum cost of fabricating a token without a payload is $c_0$
Joules and with a payload $c_{1}$ Joules.  Symbolic string tokens incur a ``per character'' cost
which we define as $\Delta c_{1}$ per character per token. For example, adding a nucleotide to
double-stranded DNA requires $2$ ATP ($1.6 \times 10^{-19}$ J) while adding an amino acid to a
protein requires $4$ ATP ($3.2 \times 10^{-19}$ J) \cite{lehninger2005}.  Apart from the per residue
per token cost, there may be other energy involved in sequestration, release and/or token transport
across a gap.  However, the key assumption is constant energy use per token. Without considering the
details as in \cite{kurana} we will denote the combination of these and any other relevant energies
as $c_e$ Joules per token. Thus, our power for the timing only channel is 
\bee
\label{eq:Ptimingonly}
{\cal P}_T = \mylambda (c_0 + c_e)
\eee
and for the timing plus payload channel, 
\bee
\label{eq:Ppayload}
{\cal P}_{T+P}
\le
\mylambda \left  ( c_1 +  c_e + \left ( \frac{\Hup(\Tmat)}{\log b}+  K \right ) \Delta c_1 \right )
\eee
where $K$ is the string length of information-laden tokens, and $b$ is the alphabet size used to
construct the strings.  For amino acids (alphabet) are used to construct proteins (strings), and in
general, monomers are used to construct oligomers.  The inequality in \equat{Ppayload} results from
the fact that $\Hup(\Tmat)$ is an upper bound on the ordering entropy $H(\Omega|\Svv,\Tmat)$ over
all possible first-passage distributions (\Thmref{Hbound}).

We also note that information could be carried {\em only} with payload (and not timing).  The power
budget would be identical to that of \equat{Ppayload} except that $\Hup(\Tmat)$ would be
replaced by $\lim_{M\rightarrow \infty} \min_{\tv} H(\Omega|\Svv,\tv)/M$. However, since
$H(\Omega|\Svv,\Tmat) \ge \min_{\tv} H(\Omega|\Svv,\tv)$, \equat{Ppayload} provides an upper bound
for the payload-only power as well.

\blankout{
\note{placeholder equation}
\bee
\label{eq:Ppayloadonly}
{\cal P}_{P}
\ge \mylambda \left  ( c_1 +  c_e + \left ( \frac{\frac{1}{\myrho} E_\ell \left [ \ell \log \ell \right ]}{\log b}+  K \right ) \Delta c_1 \right )
\eee}

\section{Results}
We can now define the capacities for the token timing (only), token timing plus
token payload and token payload (only) channels as follows:
\bee
\label{eq:CTO}
{\cal C}_T
=
\mylambda C_q(\myrho)
\eee
\bee
\label{eq:CT+P}
{\cal C}_{T+P} = \mylambda \left ( C_q(\myrho) + K \log b \right )
\eee
and
\bee
{\cal C}_P = \mylambda K \log b = {\cal C}_{T+P} - {\cal C}_T
\eee

\begin{figure}
\begin{center}
  \includegraphics[width=3.5in]{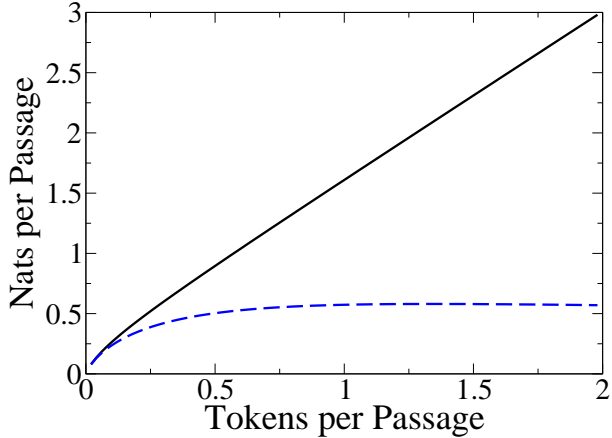}
\end{center}
\caption{Lower bound (dashed line: \Thmref{CqCtloexp}) and upper bound (solid line:
  \Thmref{Iupbound}) for the token timing channel capacity $C_t$ (in nats per passage time
  $1/\mymu$) as a function of channel load $\myrho$, the ratio of the token launch rate $\mylambda$
  to the token uptake rate $\mymu$.  }
\label{figure:upperC}
\end{figure}
In FIGURE~\ref{figure:upperC} we use \Thmref{CqCtloexp} and \Thmref{Iupbound} to plot lower and
upper bounds for ${\cal C}_T$ versus $\myrho$, a proxy for power budget, ${\cal P}$ assuming some
unit cost per token ($c_0+c_e = 1$).  {\em It is important to note} that first-passage time {\em
  variance} (jitter) produces disordered tokens. That is, the mean first-passage time is only a
measure of channel {\em latency} -- the ``propagation delay'' so to speak -- and does not itself
impact token order uncertainty.  However, for exponential first-passage the standard deviation also
happens to be the first passage time $1/\mymu$.  At small values of $\myrho$ the bounds are
tight. At larger $\myrho$ the bounds diverge and the upper bound offers the tantalizing hint that
timing channel capacity increases with increased token load $\myrho$ (see also
\cite{farsad_isit16,farsadIT16}).  Unfortunately, we have as yet been unable to find an empirical
density $f_{\Tmat}(\tv)$ which displays capacity growth similar to the upper bound and suspect that
timing capacity flattens with increasing $\myrho$ owing to a more rapidly increasing probability of
token confusion at the output.

\begin{figure}
\begin{center}
\includegraphics[width=3.5in]{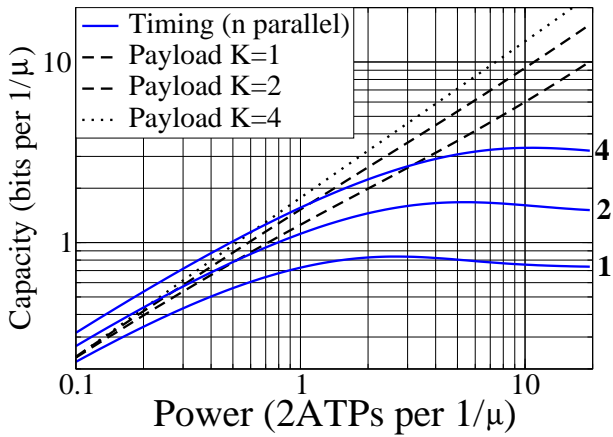}
\end{center}
\caption{Lower bounds for the capacities of the token timing ($\myrho {\cal C}_T$)
  and token timing plus token payload ($\myrho {\cal C}_{P+T}$) channels as a function of power
  budget (${\cal P}_T$ and ${\cal P}_{P+T}$) for DNA string tokens with exponential first-passage
  times (\Thmref{CqCtloexp}).  Capacity is in units bits per first-passage time $1/\mymu$.  Power is
  in units of $2$-ATP ($1.6 \times 10^{-19}$ J) per passage time $1/\mymu$ and a nucleotide residue
  is assumed to carry 2-bits of information.  Solid lines: aggregate capacity of $n=1,2,4$ separate
  (independent or parallel) token timing channels where DNA string tokens carry no information
  payload.  Dashed/Dotted lines: aggregate capacity of token timing plus token payload channels for
  DNA string tokens of different lengths, $K=1,2,4$-residue payloads.  }
\label{figure:capacities}
\end{figure}
In FIGURE~\ref{figure:capacities}, we use \Thmref{CqCtloexp} to plot lower bounds in bits per
first-passage time, $1/\mymu$, as a function of power budget ${\cal P}$ assuming DNA-based tokens.
For token timing plus token payload signaling we show plots for $K=1,2,4$ DNA-residue tokens.  For
timing-only signaling we also include plots where different identifiable tokens (different molecule
types or physically separate channels) are used ({\em i.e.}, $n=1,2,4$ parallel timing channels as
shown) for comparison with payload channels.  We have assumed costs $c_0 =\Delta c_1 = 2$ ATP.
Furthermore, we assume $c_1 = c_e = \Delta c_1$ since it seems likely that the absolute minimum
energy for token release, $c_e$, in a purely diffusive channel is probably comparable to the cost of
creating (or breaking) the covalent bond used to append a nucleotide residue. If we assume $1/\mymu
= 1$ms, then the ordinate of FIGURE~\ref{figure:capacities} is in kbit/s and the abscissa is in
units of $1.6 \times 10^{-16}$W.  If $1/\mymu = \mymu s$, (as might be the case for smaller gaps in
a nano-system) the ordinate is in $Mbit/s$ and the abscissa is in units of $1.6\times
10^{-13}$W. These data rates are many many orders of magnitude larger than the fractional bit/second
data rates previously reported for simple demonstrations of molecule communication \cite{eckford3},
and the predicted power efficiencies are startling.  Comparison of our results to \cite{eckford3}
and others would be relatively straightforward if passage time jitter for the experimental setup
were provided, although in \cite{eckford3} Avogradrian numbers of molecules were release with each alcohol
``puff'' so precise timing at the molecular level was not attempted.

Finally, it is worth noting that increasing the rate at which tokens with payload are launched will
increase the bit rate but not increase the required energy per bit.  Of particular note, at low
power, timing-only signaling provides the best rates while at higher power, inscribed matter tokens
may be preferred.  However, if it is difficult to synthesize long strings (heavily information-laden
tokens), even a single bit of information (two distinguishable species used in parallel) markedly
increases capacity.

\section{Discussion \& Conclusion}
\label{sect:discussion}
We have provided a general and fundamental mathematical framework for molecular channels and derived
some associated capacity bounds.  We now discuss the results in the context of selected prior work
and also touch upon ideas for further work suggested by the results.  We separate these into two tranches:
\begin{itemize}
\item
  {\em Engineering Implications} where we consider how molecular communication can be extended to other known communication scenarios as well where we might look for inspiration from biology that has had eons to evolve solutions.
\item
 {\em Biological Implications} where we consider how the results might impact/support known biology and suggest new avenues for investigation.
\end{itemize}

\subsection{Engineering Implications}

\noindent {\bf Capacity Bounds and Coding Methods:} Our upper bound on capacity $C_t$, the timing
capacity for identical tokens, is tight for low token load $\myrho$ but diverges for large
$\myrho$. However, no empirical distributions with rates higher than the lower bound have yet been
found. So, does the capacity of the timing-only channel truly flatten with increasing $\myrho$ as in
FIGURES~\ref{figure:capacities} and \ref{figure:upperC}, or is there a benefit to increasing the
intensity of timing-token release as suggested in \cite{farsad_isit16, farsadIT16}? In addition,
since exponential first-passage is not the worse case corruption, what {\em is} the minmax capacity
of the molecular timing channel?  Likewise, how much better than exponential might be other
first-passage densities imposed by various physical channels, and what are good codes for reliable
transmission of information over molecular channels?

While we have focused on tokens in the form of linear symbolic strings, DNA and protein sequences in
particular, what benefits might string tokens with a branched structure, exemplified by
carbohydrates, confer for sequencing and/or payload?  Should we vigorously pursue technology to
produce large payload (many residue) tokens \cite{churchDNAstorage16}, or should a pool of smaller
pre-fabricated payloads be used to deliver information?  The bunching seen in
FIGURE~\ref{figure:capacities} for payload tokens with increasing $K$ may suggest the latter when
rapid token construction is difficult.  That is, the capacity per power output does not scale
linearly in $K$ owing to the increased power required by adding more bases to tokens. This implies a
tradeoff between timing-only and increasingly larger payloads -- completely aside from the fact that
payload size, shape and composition can have an effect on transport properties.

\noindent {\bf Precise Timing, Fuzzy Timing and Concentration:} Of particular importance is
establishing a careful quantitative relationship between our finest-grain timing model and other
less temporally precise ones
\cite{bassler1999,bassler2002,fekriisit11,akyildiz_nanonet,akyildiz_diffusion,eckford1,
  Eckbook,ISIT2013_Arash, arash_wireless2013, ICC2014_Arash, eckford_press_vodka}.  To begin,
consider that our model seems to imply infinitely precise control over the release times $\Tmat$ and
infinite precision measurement of the arrival times $\Svv$.  However, release time and measurement
time imprecision are both easily incorporated into the transit time vector $\Dmat$.  Thus,
application of our model to the ``fuzzier'' release and detection times associated with
practical/real systems is straightforward.  Put another way, first passage time jitter already
imposes limits on timing precision.  Thus, so long as timing precision is significantly better than
passage time jitter, the bounds presented here will be moderately tight.  In addition, we are
hopeful that the upper bound of Theorem~\Thmref{Iupbound} will be useful for evaluating molecular
timing channel capacity for arbitrary first passage time distributions since it requires only
knowledge of the timing channel capacity coupled to average properties of the corresponding input
distribution.

Concentration is derived from considering temporal windows and counting the arrivals
within them.  Therefore, via the data processing theorem, our precise timing model {\em \bf must}
undergird all concentration-based methods which, even with perfect concentration detection, {\em
  cannot possibly exceed the capacity of the finest grain timing model presented here}. Of
particular note for the asymptotic nature of our analysis here, an individual emission schedule
$\tv$ for large $M$ is {\em exactly} a temporal emission concentration sequence as time resolution
coarsens.  That said, we have not as yet tried to show a graceful degradation toward coarse timing
concentration from precise timing.  Regardless, the results here provide crisp bounds on the
capacities derived from concentration-based models.

As a specific example, there are channel models where information is carried by the {\em number} of
molecules released and received (most recently, see \cite{farsadIT16,farsadsurvey16}). In this case,
the capacity per channel use, $C_N$, is upper bounded by $\log (M+1)$ since between $0$ and $M$
tokens can be released during a symbol interval of duration $\tau(M)$. Smaller $\tau(M)$ increases
the capacity in bits per second.  Larger $M$ increases the capacity in bits per channel use.

If we assume a fixed signaling interval $\tau(M)$ during which $m=0,1, \cdots M$ tokens are emitted,
then we can also fix the average token rate $\mylambda$ as a proxy for power. Assuming
a uniform distribution on the number of tokens sent we then have
\bee
\tau(M) = \frac{M}{2 \mylambda}
\label{eq:tauMconcentration}
\eee
since the average number of tokens released is $M/2$.

Assuming exponential first passage, the probability that all tokens arrive by $\tau(M)$ is minimized
when all tokens are launched at $t=0$. For exponential first passage and with arrival probability criterion
$1-\epsilon$ as in section~\ref{sect:channeluse} we have
\bee
\tau(M) =  - \frac{1}{\mymu} \log \left ( 1- (1-\epsilon)^{\frac{1}{M}} \right )
\label{eq:tauM}
\eee
which assures that even when $M$ tokens are emitted, they will all arrive before $\tau(M)$ with
probability $1-\epsilon$.

However, \equat{tauM} in combination with the power limit of \equat{tauMconcentration} sets 
$\mylambda$ to
\bee
\mylambda(M) = -\frac{\mymu M}{2  \log \left ( 1- (1-\epsilon)^{\frac{1}{M}} \right )}
\label{eq:lambdaMconcentration}
\eee
Then, since $C_N \le \log (M+1)$, after setting a successful channel use criterion of $1-\epsilon$ we have
\bee
\frac{C_N(M)}{\tau(M)}
\le
- \frac{\mymu \log(M+1)}{\log \left ( 1- (1-\epsilon)^{\frac{1}{M}} \right )}
\label{eq:Cnumber}
\eee
which we rewrite as
\bee
\frac{C_N(M)}{\mymu \tau(M)}
\le
- \frac{\log(M+1)}{\log \left ( 1- (1-\epsilon)^{\frac{1}{M}} \right )}
\label{eq:Cnumberrho}
\eee
with normalized power constraint
\bee
{\cal P}(M)
=
\frac{\mylambda(M)}{\mymu}
=
-\frac{M}{2  \log \left ( 1- (1-\epsilon)^{\frac{1}{M}} \right )}
\eee

However, except for small $\epsilon$, $\frac{C_N(M)}{\mymu \tau(M)}$ is not a reliable indicator of
capacity since with increased $\epsilon$, $\tau(M)$ decreases but the probability of intersymbol
interference (ISI) increases.  Since calculating the capacity of this channel with ISI is difficult,
we roughly approximate by normalizing $\frac{C_N(M)}{\mymu \tau(M)}$ by the expected number
of intervals over which a given emission burst of $M$ tokens will span (thereby corrupting potentially
them). Noting that $(1-\epsilon^z)^M$ is the probability that all tokens arrive before the end of
the $z+1^{\mbox{st}}$ interval after emission we have
\bea
\bar{z}(M) & = & \sum_{z=0}^\infty
(1 - (1 - \epsilon^z)^M) \nonumber \\
 & = & 
-\sum_{n=1}^{M} {M \choose n}  \frac{(-1)^{n}}{1-\epsilon^{n}}
\label{eq:zbar}
\eee
and we obtain
\bea
\tilde{C}_N(M) & \equiv & \frac{C_N(M)}{\bar{z}(M) \mymu \tau(M)}\nonumber \\
 & \le & 
\frac{\log \left ( M+1 \right ) }{- \bar{z}(M) \log \left ( 1- (1-\epsilon)^{\frac{1}{M}} \right )}
\IEEEeqnarraynumspace
\label{eq:numbersimplifytilde}
\eea
as an approximation to capacity for the number/concentration channel.

\begin{figure}
\begin{center}
\includegraphics[width=3.5in]{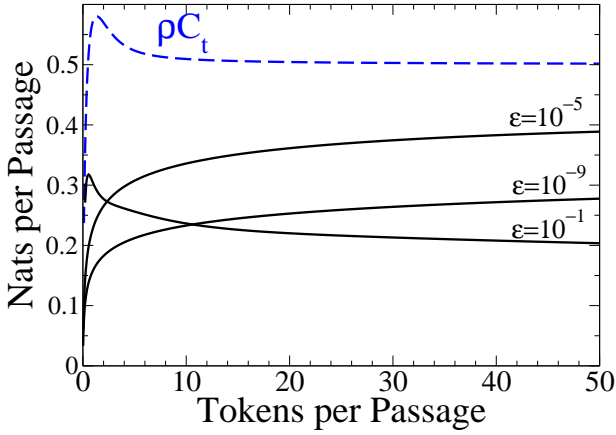}
\end{center}
\caption{Lower bound for $\rho {\cal C}_t$ versus tokens per passage (dashed line) compared to $\tilde{C}_N$
  vs. tokens per passage for different values of $\epsilon$ as shown. Token construction and emission are
  assumed unit energy for both the timing channel and the number/concentration channel.}
\label{figure:numberVtiming}
\end{figure}

In FIGURE~\ref{figure:numberVtiming} we plot the upper bound of $\tilde{C}_N(M)$ in
\equat{numbersimplifytilde} versus ${\cal P}(M)$ (i.e., parametrized in $M$) for a range of
$\epsilon$ as compared to the timing channel lower bound in FIGURE~\ref{figure:upperC}.  The timing
channel capacity lower bound is always significantly greater than $\tilde{C}_N(M)$.  Nonetheless,
the coding simplicity of the number/concentration channel could make it an attractive option.

\noindent {\bf Identifiable Tokens Without Payload:} In section~\ref{sect:payload} we mentioned the
possibility of uniquely identifying each of $M$ emitted tokens with a sequence number of length
$\log M$ bits.  We treat this scenario as distinct from ensemble timing channel coding which
resolves residual orderering ambiguity (see section \ref{sect:payload}) because if the tokens are
individually identifiable, the potential emission schedules are not constrained to ensemble timing channel
coding.  Thus, the $M$ identifiable tokens constitute $M$ parallel single-token timing channels, which
for exponential first passage have aggregate capacity $M \log (1 + \frac{M}{\myrho e})$.

However, $\myrho$ is limited by the power budget ${\cal P}$ (in units of energy per passage, $1/\mymu$)
\bee
\myrho  {\log M}
\le
{\cal P}
\label{eq:identified}
\eee
because each token requires $\log M$ bits of sequencing information energy.  Following \ref{sect:channeluse}
we have $\mylambda \tau(M)= M$ so the capacity in nats per passage time is
\bee
C
=
\myrho  \log \left  ( 1 + \frac{M}{\myrho e} \right )
\le
\myrho  \log \left  ( 1 + \frac{e^{\frac{\cal  P}{\myrho}}}{\myrho e} \right )
\label{eq:identifiedC}
\eee
the last inequality owed to \equat{identified}.  However, in the limit of $M \rightarrow \infty$ we
have $\myrho \rightarrow 0$ so we have
\bee
\lim_{\myrho \rightarrow 0} C
=
{\cal P}
\eee
in units of nats per passage time (and assuming unit per-bit cost of the token identfier string).
Thus, the identifiable token timing channel capacity exceeds the identical token timing
channel lower bound with increasing power budget and scales linearly in power as does the timing
channel upper bound (see FIGURE~\ref{figure:upperC}) -- while still lying below it.

\noindent {\bf Token Corruption and Receptor Noise:} Certain channel properties we have so far
ignored must also be studied, such as the potential for lost or corrupted tokens and potential
binding noise at receptor sites.  However, as previously stated, token erasure (tokens which do not
arrive) or payload token corruption (tokens which are altered in passage) or receptor noise (tokens
bind stochastically to the receptor) can only decrease capacity (via the data processing theorem).
Thus, the results here provide upper bounds.  Nonetheless it is worth considering how the analytic
machinery developed might be modified to take into account such impediments.

First, consider alteration of payload-carrying tokens {\em en route}.  If the corruption is
i.i.d. for each token, then the usual error correcting coding methods can be applied individually,
or to the token ensemble.  The resulting overall channel capacity will be degraded by the coding
overhead necessary to preserve payload message integrity (including the sequencing headers).

Then consider token erasure where a token never arrives (and is assume to not arrive in a later
signaling interval).  Since each signaling interval uses $M$ tokens, we will know whether tokens get
``lost'' in transit and can arbitrarily assign a faux arrival time to such tokens.  However, the
problem this poses for our analysis is two-fold.  First, tokens released later in the
signaling interval are more likely to be lost which implies that the first passage density is not
identical for each token.  Second, the first passage density for each token would then contain a
singularity equal to the probability of loss, violating one of our key assumptions and making
hypersymmetric probability density folding arguments invalid.  That said, an erasure channel
approach where tokens were deleted randomly from the output could be pursued and owing to its
i.i.d. nature (with respect to which tokens were erased) would likely provide a worst case
scenario, since the information associated with erasures being more likely to be derived from a
later release would be absent.

Finally, we have previously mentioned that a token may ``arrive'' multiple times owing to receptor
binding kinetics.  It was previously shown that given the first binding (first-passage) time the
information content of subsequent bindings by the same token is nil \cite{eckford1}. In addition, as
shown in section~\ref{sect:channeluse}, information-theoretically patent channel use requires that
tokens from a given emission interval be eventually cleared at the receiver.  Otherwise, lingering
tokens can interfere with subsequent emission intervals.  

So, one could imagine that the rebinding process results in a characteristic finite-mean ``burst''
of arrivals associated with a given token which could perhaps be resolved into a single
first-arrival time estimate -- effectively adding more jitter to the first-passage time $D$.  If so,
our model applies directly with appropriate modification.  However, we have not attempted to analyze
this scenario nor quantify the associated estimation noise.  Thus, our results are most
appropriately applied to systems where ligands bind tightly or where tokens are removed with high
probability after first capture/detection. The addition of noisy ligand binding can only depress the
capacity bounds presented here and we leave the question of exactly how much to future work.

\noindent {\bf Interference and Multiple Users:} Multi-user communication in a molecular setting is
a critical question, and a better understanding of the single-user channel will certainly help with
multi-user studies where transmissions interfere. There is some work to inspire an
information-theoretic edifice \cite{moewin16} similar to how the current work builds on
\cite{bits-Qs}, but the multi-user molecular signaling problem has not yet been rigorously
considered.  Of particular interest would be a version of MIMO since FIGURE~\ref{figure:capacities}
shows capacity benefits to parallel channels.  One could imagine apposed arrays of emitters and
receivers which could be engineered to collaborate to encode and decode information in a variety of
ways, from parallel non-interfering channels to grossly interfering channels where joint/distributed
coding might be employed.  One could even imagine channels with chemically reactive species in which
emitted tokens elicited spatially structured propagation of detectable reaction products
\cite{chemwaves1,chemwaves2,chemwaves3}.

\noindent {\bf Other Applications:} It is interesting to note that although our work is couched in
terms of molecular communication, the notion of token inscription applies to any system where
discrete emissions experience random transport delay between sender and receiver.  The most obvious
example is the Internet where packets experience variable delay and may arrive out of order.  Our
results provide crisp bounds on the amount of sequencing overhead necessary for proper message
reconstruction and even suggest that (at least for low payload packets traveling over independent
routes) timing information could be an interesting adjunct to payload, depending upon the amount of
timing jitter between the source and destination.

There is also the potential for cross pollination from biology to communication systems. As a simple
example, there may be some selective benefit to hiding information from competitors.  So, perhaps
biological systems, where signaling chemicals are often detectable by other organisms, convey
secrets over molecular communication channels in ways that can be mimicked in engineered systems.
An obvious application, biosteganography \cite{biosteganography16}, comes to mind in the context of
tokens with payloads, although in such schemes timing plays no role as yet.

\subsection{Biological Implications}

Clearly, the natural world clearly offers a dizzying array of processes and phenomena through which
the same and different tasks, communication or otherwise, might be accomplished (see, for example,
\cite{Joussineau_filopod,GorYan06,GurBar08,WanVer10,purcell_chemophys,Bassler09}).  It is no wonder
therefore that communication theorists have plied their trade heavily in this scientific domain (for
a relatively recent review, see \cite{todd_ITspecial10}).  Identifying the underlying mechanisms
(signaling modality, signaling agent, signal transport, and so on) as well as the molecules and
structures implementing the mechanisms is no small undertaking. Consequently, experimental
biologists use a combination of prior knowledge and what can only be called instinct to choose those
systems on which to expend effort. Guidance may be sought from evolutionary developmental biology --
a field that compares the developmental processes of different organisms to determine their
ancestral relationship and to discover how developmental processes evolved. Insights may be gained
by using statistical machine learning techniques to analyze heterogeneous data such as the
biomedical literature and the output of so-called ``omics'' technologies -- genomics (genes,
regulatory, and non-coding sequences), transcriptomics (RNA and gene expression), proteomics
(protein expression), metabolomics (metabolites and metabolic networks), pharmacogenomics (how
genetics affects hosts' responses to drugs), physiomics (physiological dynamics and functions of
whole organisms), and so on.

Frequently, the application of communication theory to biology starts by selecting a candidate
system whose components and operations have been already elucidated to varying degrees using methods
in the experimental and/or computational biology toolbox \cite{HH4,Long09} and then applying
communication and/or information theoretic methods
\cite{todd_ITspecial10,wojtek,donjohnson10,emery04,Bassler09,MouDia16,MouKav16,UdaKur16}.  However,
we believe that communication theory in general and information theory in particular are not mere
system analysis tools for biology but new lenses on the natural world \cite{MianRose11}.  Here, we
have sought to demonstrate the potential of {\em communication theory as an organizing principle for
  biology}. That is, given energy constraints and some general physics of a problem, an
information-theoretic treatment can be used to provide outer bounds on information transfer in a
{\em mechanism-blind} manner.  Thus, rather than simply elucidating and quantifying known biology,
communication theory can winnow the plethora of possibilities (or even suggest new ones) amenable to
experimental and computational pursuit. Likewise, general application of communication-theoretic
principles to biology affords a new set of application areas for communication theorists.

Examples of the implications of our main communication theoretic results are as
follows:
\begin{itemize}
\item
{\em We have derived a model and methodology for determining the amount of information a system
  using chemical signaling can convey under given power constraints.}  Do (or How do) biological
systems achieve this extremely -- even outrageously -- low value?
\item
{\em Using tokens with large payloads can be very efficient.}  Is one example of this the
transmission of hereditary material such as a genome over evolutionary time scales (periods spanning
the history of groups and species)?
\item
{\em If it is difficult to synthesize long strings (information-laden tokens), even a single bit of
  information (two distinguishable species) increases capacity.}  {Is the transmission of hormones,
  semiochemicals and other small molecules over developmental time scales (periods spanning the life
  of individuals) indicative of some efficiency tradeoff?}
\item
{\em Although the production rate of tokens can be increased, the channel capacity might not
  increase if tokens are emitted at intervals smaller than the variability of arrival time.}  {Does
  the material through which tokens travel hold the key to addressing this and the aforementioned
  questions and problems about engineered systems -- particularly the issues of concentration versus
  timing, token corruption and interference and multiple users? In biological systems,
  discrete particles propagate through a cornucopia of substances en route from the source to the
  target: solids, liquids, gases or plasmas in the biosphere, lithosphere, atmosphere, hydrosphere,
  interstellar space or intergalactic space -- for instance cytoplasm, nucleoplasm, mitochondrial
  matrix, extracellular matrix, extracellular polymeric substances, blood, lymph, phloem, xylem,
  bones, soil, rocks, air, water, steam, ice, and molecular clouds. Clearly, the ``propagation
  delay'' (jitter) experienced by a particular category and type of discrete particle is a function
  of the intrinsic physicochemical properties of the material. The standard deviation of arrival
  times will depend also on environmental factors such as temperature, pH, pressure and light.}
\end{itemize}

Our results could guide studies aimed at answering three key questions biologists ask of a living
organism: How does it work?, How is it built?, and How did it get that way? \cite{Bre12}. This is
because our models of token timing, payload and timing+payload channels are inspired, at least in
part, by fundamental ``systems'' problems about the dynamic and reciprocal relationship(s) between
individuals in multicellular systems -- whether microrganisms in communities or cells in metazoan
tissues. Inscribed matter communication is a keystone of how individuals learn what to become or to
be by a combination of internal and external cues and how, in turn, they teach others when to change
or remain the same. Thus, the seemingly esoteric theoretical studies of channel capacities described
here and discussed further in our companion paper might help pave the way to elucidating the origins
(evolutionary developmental biology), generation (embryogenesis, and morphogenesis), maintenance
(homeostasis, tolerance, and resilience), subversion (infectious and chronic diseases such as cancer
and immune disorders), and decline (aging) of complex biological form and function
\cite{MianRose11}.

A key virtue of the token timing model is its implicit acknowledgment of the importance of the
``transmission medium'' in the spatial gap between sender and receiver and through which tokens
move. Consider molecular inscribed matter communication from the microscopic to the macroscopic
levels: within and between cells in tissues and organisms in (agro)ecosystems. Since the presence of
obstacles influences the mobility of discrete particles through a material, a crowded environment
will increase the mean arrival time relative to unhindered diffusion but should have less of an
impact on the mean emission time. Decades of laboratory {\em in vitro} studies have promulgated the
view of the cellular interior as a place proteins, nucleic acids, carbohydrates, lipids and other
molecules exist as highly purified entities that act in isolation, diffusing more or less freely
until they find their cognate binding partner. In its natural milieu however, a molecule lives and
operates in an extremely structured, complex and confining environment: one where it is surrounded
by other molecules of the same or different chemical nature, the bystanders in the crowd having
positive or negative effects on its mobility, biochemistry and cell biology
\cite{MitCho15,KuzZas15}. Widening the spatial and temporal horizon, semiochemicals diffusing
through soil, water, and air mediate the complex ways crops, livestock, and microbes interact with
one another.

Whether molecule release and capture occurs among organisms in the above- and below-ground
environments or between cells in the tissue microenvironment, the basic physics is similar.  For
this reason we feel that our fundamental treatment of {\imc} communication presented here could help
guide biological understanding and experimentation.

\section*{Acknowledgments}
Profound thanks are owed to A. Eckford, N. Farsad, S. Verd\'u and V. Poor for useful discussions and
guidance.  We are also extremely grateful to the editorial staff and the raft of especially careful
and helpful anonymous reviewers. This work was supported in part by NSF Grant CDI-0835592.

\bibliographystyle{unsrt}
\bibliography{MERGE11,sm_collection,molecule-rose,biologicalcommunication,biologicalcommunicationCHRIS,mian2,NIS,RosePriorStandalone,Arash,citations,sm_collection}
\end{document}